\documentclass[preprint2]{aastex}

\shorttitle{BAT Galactic survey}
\shortauthors{Voss \& Ajello}

\begin{document}


\title{{\it Swift}-BAT Survey of Galactic Sources: 
Catalog and Properties of the populations}

\author{R. Voss\altaffilmark{1}}
\affil{Max Planck Institut f\"{u}r Extraterrestrische Physik, P.O. Box 1603, 85740, Garching, Germany}
\email{rvoss@mpe.mpg.de}
\and
\author{M. Ajello}
\affil{SLAC National Accelerator Laboratory and Kavli 
Institute for Particle
Astrophysics and Cosmology, 2575 Sand Hill Rd, Menlo Park, CA 94304, USA}
\altaffiltext{1}{Excellence Cluster Universe, Technische Universit\"at M\"unchen, Boltzmannstr.
2, D-85748, Garching, Germany}

%
%

\begin{abstract}
We study the populations of X-ray sources in the Milky Way in the
15-55 keV band using a deep survey with the \textit{BAT} instrument 
aboard the \textit{Swift} observatory. We present the $\log$N-$\log$S
distributions of the various source types and we analyze their variability
and spectra. For the low-mass X-ray binaries (LMXBs) and the high-mass
X-ray binaries (HMXBs) we derive the luminosity functions to a limiting
luminosity of $L_X\sim7 \times10^{34}$ erg s$^{-1}$. Our results confirm the
previously found flattening of the LMXB luminosity function below a
luminosity of $L_X\sim10^{37}$ erg s$^{-1}$. The luminosity function of
the HMXBs is found to be significantly flatter in the 15-55 keV band than
in the 2-10 keV band. From the luminosity functions we estimate the
ratios of the hard X-ray luminosity from HMXBs to the star-formation rate,
and the LMXB luminosity to the stellar mass. We use these to estimate the
X-ray emissivity in the local universe from X-ray binaries and show that
it constitutes only a small fraction of the hard X-ray background.

\end{abstract}

\keywords{Galaxy: stellar content -- X-rays: binaries -- X-rays: stars}

%
%

\section{Introduction}
\label{intro}
Large galaxies typically contains hundreds of bright ($>10^{36}$ erg s$^{-1}$)
X-ray sources, of which the majority are high-mass X-ray binaries (HMXBs) or
low-mass X-ray binaries (LMXBs). The possibilities of studying them in
external galaxies with \textit{XMM-Newton} and \textit{Chandra} have
sparked interest in studying the populations of these sources in galaxies.
The Milky Way provides a useful reference for such studies. 
Besides a considerable population of HMXBs in the Magellanic clouds \citep[e.g.][]{Liu2005},
and the detection of a few individual sources in other nearby galaxies
\citep[e.g.][]{Pietsch2006}, 
the Milky Way is the only galaxy in which it is currently,
or in the near future, possible to obtain information on a high fraction
of the X-ray
binaries from measurements in other wavebands than the X-rays. 
It is also
possible to measure X-ray sources in the Milky Way at much lower
luminosities than in external galaxies. The X-ray 
source populations in the Milky Way can therefore provide us with 
unique observational constraints.

However, the analysis of the population of X-ray sources in the Galaxy suffer
from several problems. The Galaxy has a large angular size and the distances
to many of the sources are not known. The population of sources is mixed and 
from X-rays alone it is not always possible to distinguish a weak nearby
source from a more distant bright source. Focusing telescopes have small
fields of view and are therefore not suited for such studies.
\citet{Grimm2002} used the All-Sky Monitor (ASM) of the RXTE observatory
to study the populations of X-ray sources in the 2-10 keV band, and
constrained the luminosity functions of X-ray binaries with
luminosities $\gtrsim10^{36}$ erg s$^{-1}$. They
found that the differential luminosity function of the HMXBs could be 
approximated by a single power-law with a slope of $\Gamma\simeq-1.6$,
whereas the luminosity function of LMXBs was more complicated with a
steep slope at high luminosities and a shallow slope at luminosities
below $10^{37}$ erg s$^{-1}$. With \textit{XMM-Newton} and \textit{Chandra}
the investigations were extended to also cover nearby galaxies. 
\citet{Grimm2003} found that the luminosity function of HMXBs in a sample 
of star-forming galaxies showed no evidence of variation, and was
consistent with the results from the Milky Way slope of $\Gamma\simeq-1.6$,
with a possible cut-off at very high luminosities, a few$\times10^{40}$ 
erg s$^{-1}$. Investigations of the LMXB populations in old stellar
environments also found results consistent with the Milky Way results,
with a steep slope at high luminosities and a shallower slope at lower
luminosites \citep[e.g.][]{Kim2004,Voss-cena,Voss-m31,Kim2009}. 
However, the exact shape remains controversial. 
\citet{Gilfanov2004} combined results from nearby galaxies with those
of the Milky Way and found a power-law slope of $\Gamma\simeq-1.0$
at luminosities below $10^{37}$ erg s$^{-1}$, and a slope of $\Gamma\simeq-1.8$
above this limit, breaking to an even steeper slope at luminosities
above $5\times10^{38}$ erg s$^{-1}$. \citet{Kim2004} studied a larger
sample of galaxies, and confirmed the slope of $\Gamma\simeq-1.8$ at
luminosities above a few times $10^{37}$ erg s$^{-1}$, and 
while a single power-law fit was acceptable, the
fit did improve when a break to a steeper slope at high luminosities was
included. Studies of the bulge of M31 and the early type galaxy
Centaurus A (Cen A, NGC 5128) showed a clear break at low luminosities
$\sim10^{37}$ erg s$^{-1}$ confirming the low-luminosity slope of
$\Gamma\simeq-1.0$ \citep{Voss-cena, Voss-m31, Voss-cena2}.
While initial studies of the elliptical galaxies NGC 3379 and NGC 4278
\citep{Kim2006} did not show any evidence of this break, 
deeper observations indicate some flattening towards low luminosities
\citep{Kim2009}. Finally, recent results \citep{Voss-m31,Woodley2008,
Voss-cena2,Kim2009} show that there is a difference between the luminosity
functions of LMXBs in globular clusters and those outside, with a
dearth of low-luminosity sources in globular clusters.\\

The \textit{RXTE ASM}, \textit{Chandra} and \textit{XMM-Newton} 
observatories are only detecting photons below $\sim$10 keV. 
However, many X-ray binaries emit a significant fraction of their
energy in harder X-rays. Incomplete knowledge of the different
X-ray states, and the time individual sources stay in these states,
makes it difficult to extrapolate the observations below 10 keV to
wider bands. Furthermore these telescopes are biased against objects
with high absorbing column densities $>10^{22}$cm$^{-2}$ 
\citep[see e.g. Figure 3 in][]{Ajello09rev}, such as
the very absorbed HMXBs recently discovered with \textit{INTEGRAL}
\citep[e.g.][]{Walter2006}.
The first observatory useful for population studies
of the Milky Way in hard X-rays ($>$10 keV) was \textit{INTEGRAL} 
with its coded-mask telescopes. This capability was used by 
\citet{Lutovinov2005} to study the spectra and 
spatial distribution of the Galactic population of HMXBs. A
similar study of the LMXBs, including the luminosity function, was
carried out by \citet{Revnivtsev2008}, but this study was limited
to the bulge LMXBs.\\

In this paper we extend the study of the populations of Galactic
X-ray sources in the hard X-rays, utilizing data obtained by the
\textit{Burst Alert Telescope} \citep[BAT;][]{barthelmy05}, on
board the \textit{Swift} satellite \citep{gehrels04}. We follow
the approach of \citet{Grimm2003}, compiling a catalogue of sources
based on previously published identifications. These are then analyzed
taking into account the limits of the identification procedures.

%
%
\section{The BAT X-ray Survey}
\label{subsec:batsurvey}

The \textit{BAT} represents
a major improvement in sensitivity for imaging of the hard X-ray sky.
BAT is a coded mask, wide field of view, telescope 
sensitive in the 15--200\,keV energy range.
BAT's main purpose is to locate Gamma-Ray Bursts (GRBs).
While chasing new GRBs, 
BAT surveys the hard X-ray sky with an unprecedented sensitivity.
Thanks to its wide FOV and its pointing strategy, 
BAT monitors continuosly up to 80\% of the sky every day. Therefore
the light-curves of all sources are sampled regularly in a manner
similar to the \textit{RXTE ASM}. Many X-ray sources are highly
variable on a variety of timescales, and therefore regular sampling 
is important for
deriving the average properties of objects, as opposed to pointed
observations that are useful for deriving the physical properties of
objects at specific times.
Results of the \textit{BAT} survey \citep{markwardt05,Ajello2008} 
show that \textit{BAT} reaches a sensitivity of $\sim$1\,mCrab in 
1\,Ms of exposure except near bright sources or very crowded
fields, where the high backgrounds can worsen the sensitivity by a
factor of $\sim2$. Given its sensitivity and the large exposure already 
accumulated in the whole sky, \textit{BAT} poses itself as an excellent 
instrument for studying the Galactic source populations. 

%
%
\subsection{Data Processing}
For the analysis presented here, we used all the available BAT data
taken from January 2005 to March 2007.
The chosen energy range for the all-sky analysis is 15--55\,keV. The lower
limit is dictated by the energy threshold of the detectors. The upper limit
was chosen as to avoid the presence of strong background lines which
could worsen the overall sensitivity. 
Data were processed using standard Swift software contained in the
HEASOFT 6.3.2 distribution.
Data screening was performed according to \cite{Ajello2008}.
We recall here the main steps. Data are filtered according to the 
stability of the pointing, the BAT  array rate ($\leq$ 18000\,Hz),
the distance to the South Atlantic Anomaly, the goodness of the 
fit to the BAT array background ($\chi_{red}<$1.5) 
and the presence of known sources
at the correct position in the FOV. Only those data which fullfill
these criteria are used for the analysis. The main difference from 
\cite{Ajello2008} is that we integrate over energy in the 15-55\,keV band
instead that in the 14--170\,keV band.
The all-sky image is obtained as  
the weighted average of all the shorter (per-pointing) observations.
For this analysis, we consider only the sky region along the
Galactic Plane whose absolute Galactic latitude is less than 20$^{\circ}$.
The average exposure in the Galactic region is  2.6\,Ms, being 1.3\,Ms 
and 4.1\,Ms the minimum and maximum exposure times respectively. 
The final  image shows
a Gaussian normal noise and we identified source candidates
as excesses above  the 4.8\,$\sigma$ level.
Above this threshold, we detected 228 objects.
Considering that the all-sky image has a pixel size of 
$8\times8$ arcmin for a total of 2.25 million pixels,
we expect $\sim$1.8 spurious detection above the  4.8\,$\sigma$ threshold
($\leq$1\,\% of the total number of excesses). 

All the candidates are fit with the BAT point spread function 
(using the standard BAT
tool {\it batcelldetect}) to derive the best source position.
The sources found in this way are all those whose averaged emission
is above the sensitivity limit of our survey ($\sim1-2\times10^{-11}$erg cm$^{-2}$ s$^{-1}$in the 15-55 keV band, depending on the local exposure and background) 
at the position of the source.
Fast transients, which are detected in the per-pointing analysis only,
are not discussed here and their study will be left to a future
publication.


%
%
\subsection{Source Identification}

We used high-energy catalogs in order to identify BAT sources.
Identification was in most cases a straightforward process, since
the cross-corralation of  BAT objects with 
the ROSAT All-Sky Survey Bright Source Catalogue \citep{voges99}
provides an easy and solid way to identify a large fraction ($\sim$70\%) 
of them \citep{Ajello2008}.
Most of the uncorrelated sources are not present in the ROSAT survey
because of absorption (either along the line of sight or intrinsic 
to the source).
However, given the very large exposure INTEGRAL accumulated along
the Galactic plane, most of the remaining sources were identified
using the Third IBIS Catalog \citep{bird07} and the INTEGRAL
all-sky catalog \citep{krivonos07}.

\begin{figure}
\begin{center}
\resizebox{\hsize}{!}{\includegraphics[angle=0]{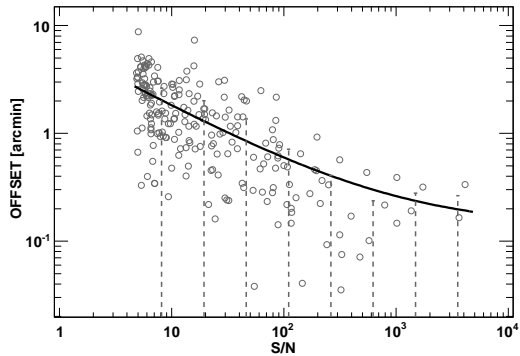}}
\caption{Offset from catalog position for the sources reported in
Table~\ref{tab:cat} as a function of S/N (open circles). 
The solid  line represents
 the best fit to the data (see Eq.~\ref{eq:off}) and gives the
mean offset vs. S/N.
The dashed lines show the standard deviation of the offset
distribution in several bins of S/N.
\label{fig:offset}}
\end{center}
\end{figure}

\begin{figure}
\begin{center}
\resizebox{\hsize}{!}{\includegraphics[angle=0]{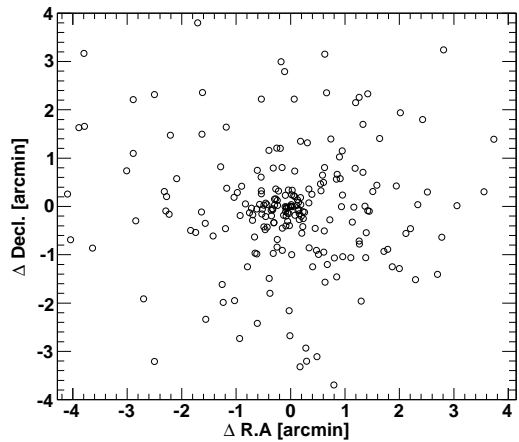}}
\caption{Difference in right ascension and declination between
the positions of the BAT sources and the position of their
 optical counterparts.
\label{fig:deltapos}}
\end{center}
\end{figure}

We report in Fig.~\ref{fig:offset}, the offset of the BAT sources
from the catalogged counterpart as a function of S/N.
We determine that the mean 
offset varies with significance according to
\begin{equation}
\label{eq:off}
\textup{OFFSET} = (6.1\pm1.5)\times (S/N)^{-0.56(\pm0.20)} + 0.13 \ \ \ \ \textup{(arcmin),}
\end{equation}

where the constant of 0.13\arcmin\ is due to a systematic misalignment
of the boresight which causes the systematic offset of the
brightest sources \citep[see also ][]{tueller10}. 
At the detection threshold of 4.8\,$\sigma$ the average
offset is $\sim$2.6\arcmin\ .
Moreover, Fig.~\ref{fig:offset} shows  the standard
 deviation of the data for different logaritmic bins of source significance. 
This is found to be always less than 2.5\arcmin\ .
Moreover, Figure~\ref{fig:deltapos} shows the difference in the celestial
coordinates
between the  position of the BAT sources
and the position of the optical counterpart.
In both directions (e.g. right ascension and declination) the distributions
are centered in zero and exhibit a similar standard deviation of 1.5\arcmin\ .
All these 
results confirm the good position accuracy of BAT even in crowded regions
as the Galactic plane.

%
%
\section{Catalog}
\label{sec:cat}
In Table~\ref{tab:cat} we report the coordinates, fluxes and other
details of the 228 detected sources.
Most of the objects have both an identification
in other X-ray band and in the optical. In a few cases, the optical
classification is still uncertain or unknown.
Only in 5 cases we do not have a secure identification for the BAT
object, and 12 further sources do have
counterparts, but have unidentified object types. For 4 of the 5
sources without identification we have listed tentative IDs in
Table \ref{tab:uid}.
The fluxes quoted in Table~\ref{tab:cat} are time-averaged fluxes over the whole
data set in the 15--55\,keV energy band. Conversion from count rate
to flux was performed adopting a Crab Nebula spectrum  of the 
form $dN/dE=10.17\ E^{-2.15}$. Position uncertainty for the 
BAT objects can be derived, as a function of significance, using 
Equation~\ref{eq:off}. We derive that the average location accuracy
for a 5, 10 and 20$\sigma$ source is 2.6\arcmin\ , 1.8\arcmin\
 and 1.2\arcmin\ .
For comparison, the location accuracies reported for INTEGRAL-IBIS for the 
same significances are 
2.1\arcmin\ , 1.5\arcmin and 0.8\arcmin\ respectively \citep{krivonos07}.
The better location accuracy of INTEGRAL-IBIS is not surprising in
view of the fact that the IBIS point spread function is sharper 
than the BAT one \citep[12\arcmin\ versus 22\arcmin\
full width at half maximum, see ][]{bird06,barthelmy05}.

\begin{figure*}
\begin{center}
\resizebox{0.7\hsize}{!}{\includegraphics[angle=0]{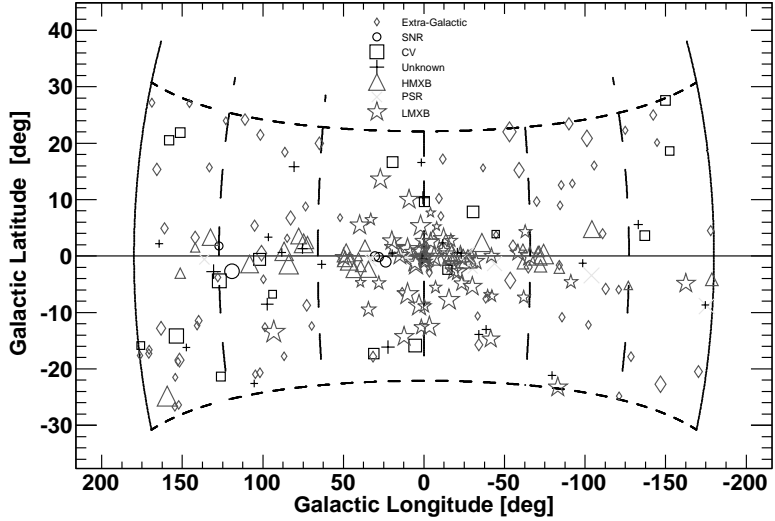}}
\caption{AITOFF projection of the distribution of sources on the sky, 
divided by source type. The size of the symbols is proportional to the 
source flux.}
\label{fig:galplot}
\end{center}
\end{figure*}

Many of the X-ray binaries have known distances, albeit with 
large uncertainties, and the
catalogue includes the approximate distances to HMXBs taken from \citet{Liu2006}
and to LMXBs from \citet{Liu2007}. In Table \ref{tab:objtyp} we give the
numbers of different identified source types, and in Figure \ref{fig:galplot}
the distribution of the source types on the sky is shown.

\begin{figure*}
\begin{center}
\resizebox{0.9\hsize}{!}{\includegraphics[angle=0]{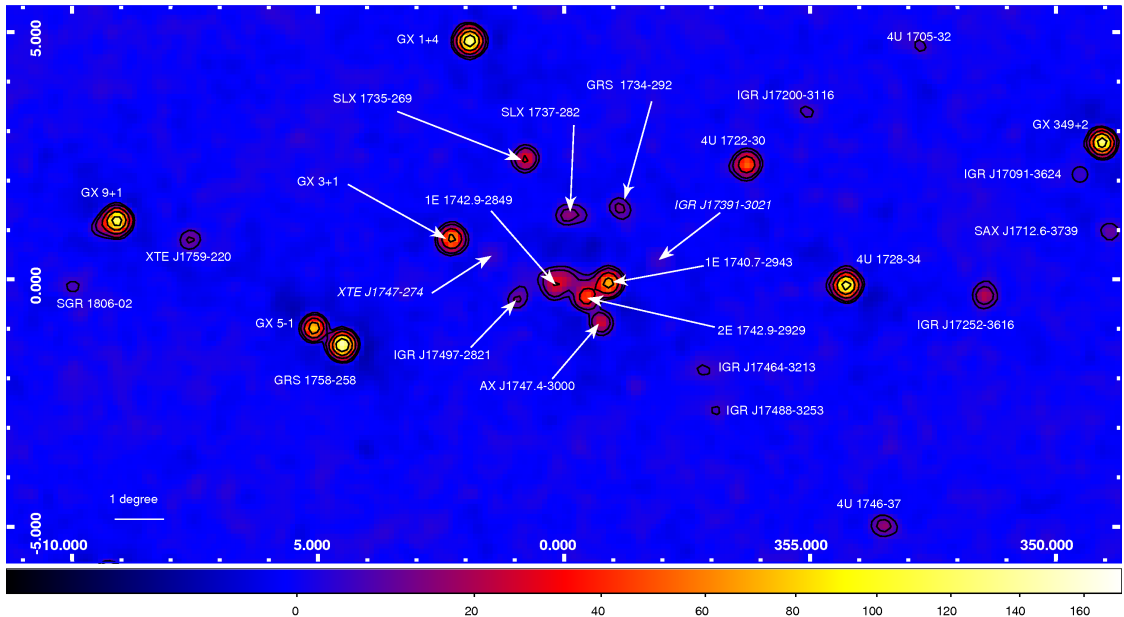}}
\caption{Significance image of the Galactic Center 
region as seen by Swift/BAT in the 15--55\,keV band. Black contours
denote leveles of S/N which start from S/N = 4.8 and stop at S/N = 100
with a multiplicative factor of 2. The x and y axes show the Galactic
longitude and latitude respectively.}
\label{fig:gc}
\end{center}
\end{figure*}

Figure~\ref{fig:gc} shows the inner 20$^{\circ}\times$10$^{\circ}$ region
around the Galactic center where BAT detects more
than 30 sources. Particularly, 
when looking at the Galactic center,
the similarity of the BAT and the INTEGRAL images is apparent
\citep{revnivtsev04a,bird06} although  BAT is unable to resolve all
the sources in this complex region. Two of the sources reported
in the map  are not part of this work because their significance,
when integrated over the 2 years of the survey,
is lower than 4.8\,$\sigma$. Indeed, they are transient sources
which are detected by BAT only during their outburst episodes.
One source is XTE J1747-274 which is a neutron star LMXB
which was very active particularly
in March-April 2005 \citep[see e.g.][and references therein]{zhang09}.
The other source, IGR J17391-3021, is a supergiant fast X-ray
transient caracterized by very short intense bursts lasting
on the order of hours \cite[e.g.][]{smith06}. This source
was particularly active in BAT during 2006.

%
%

%
%
\section{Spectral Properties}
\label{sec:spec}
For each object in our survey we extracted a 15--195\,keV spectrum  
with the method 
described in \cite{Ajello07b}. Here we recall the main steps: for a given
source, we extract a spectrum from each observation where the source is in the
field of view. These spectra are corrected for residual background 
contamination
and for vignetting; the 
per-pointing spectra are then (weighted) averaged to produce the final source
spectrum. Thus, the final spectrum represents the average source emission 
over the time-span considered here. The accuracy of these
spectra is discussed in details in \cite{Ajello09rev}.

The average spectral  properties of the sample can be studied by means
of hardness ratios (HR) which are defined as normalized differences
between the background-subtracted count-rates in the
 soft band (S, 14--22\,keV), in the medium band (M, 22--50\,keV)
and in the hard band (H, 50--195\,keV).

\begin{figure}
\begin{center}
\resizebox{\hsize}{!}{\includegraphics[angle=0]{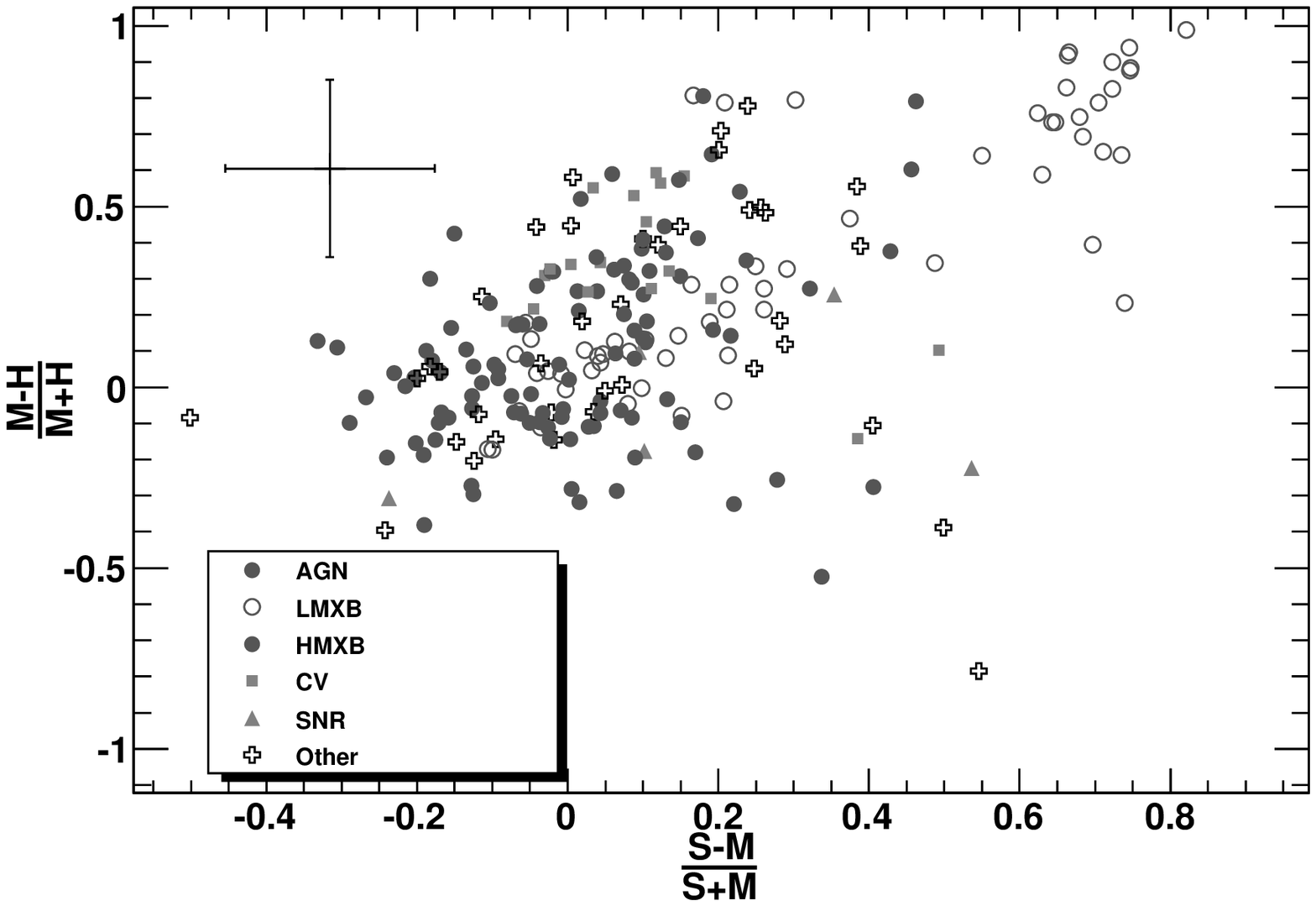}}
\caption{Hardness ratio plot of the BAT Galactic sample.
In the upper left corner the typical $\pm1\sigma$
error for a 5\,$\sigma$ source is shown.}
\label{fig:HR}
\end{center}
\end{figure}

\begin{figure*} 
    \begin{center}
     \begin{tabular}{cc}
   \includegraphics[scale=0.4]{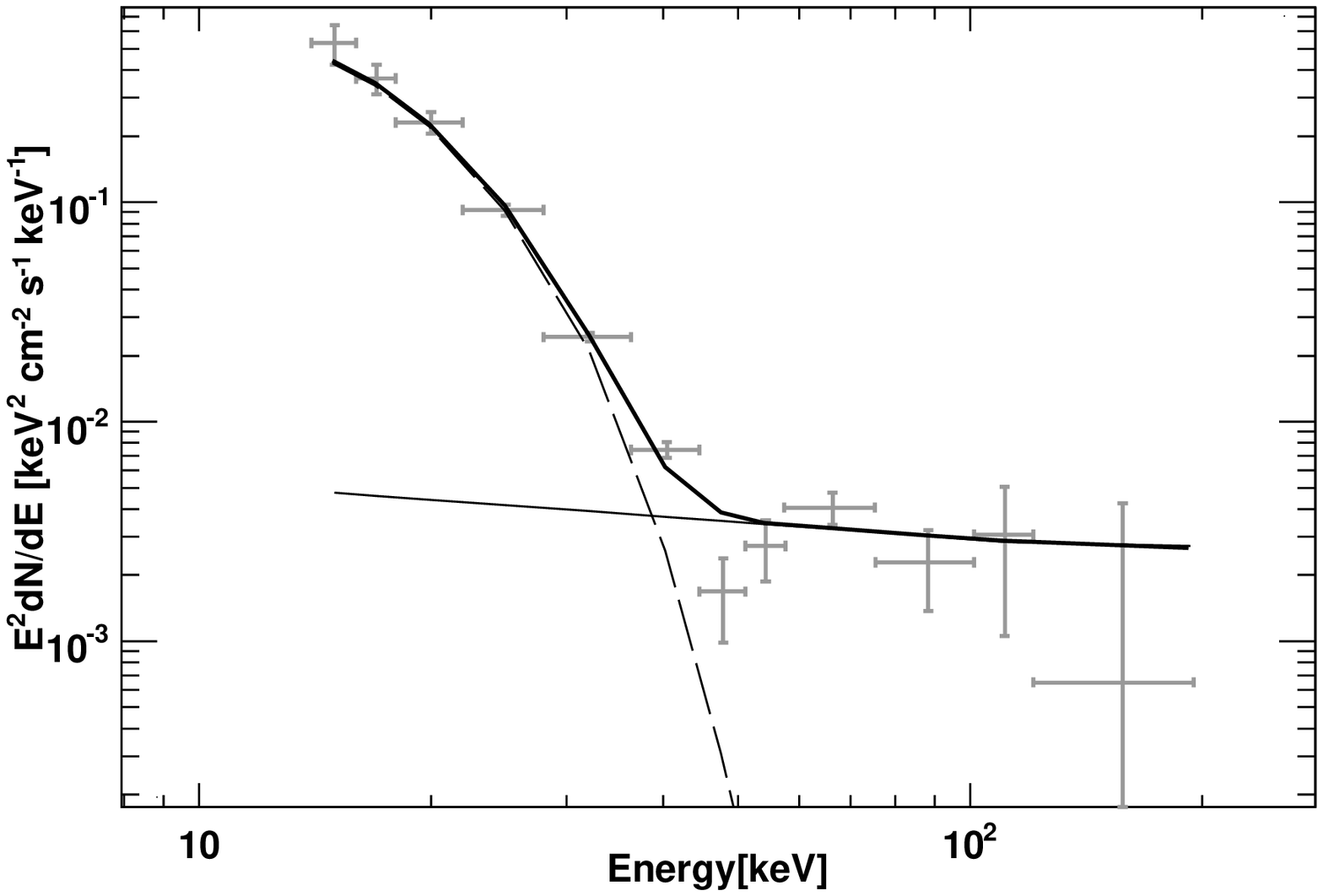} &
    \includegraphics[scale=0.4]{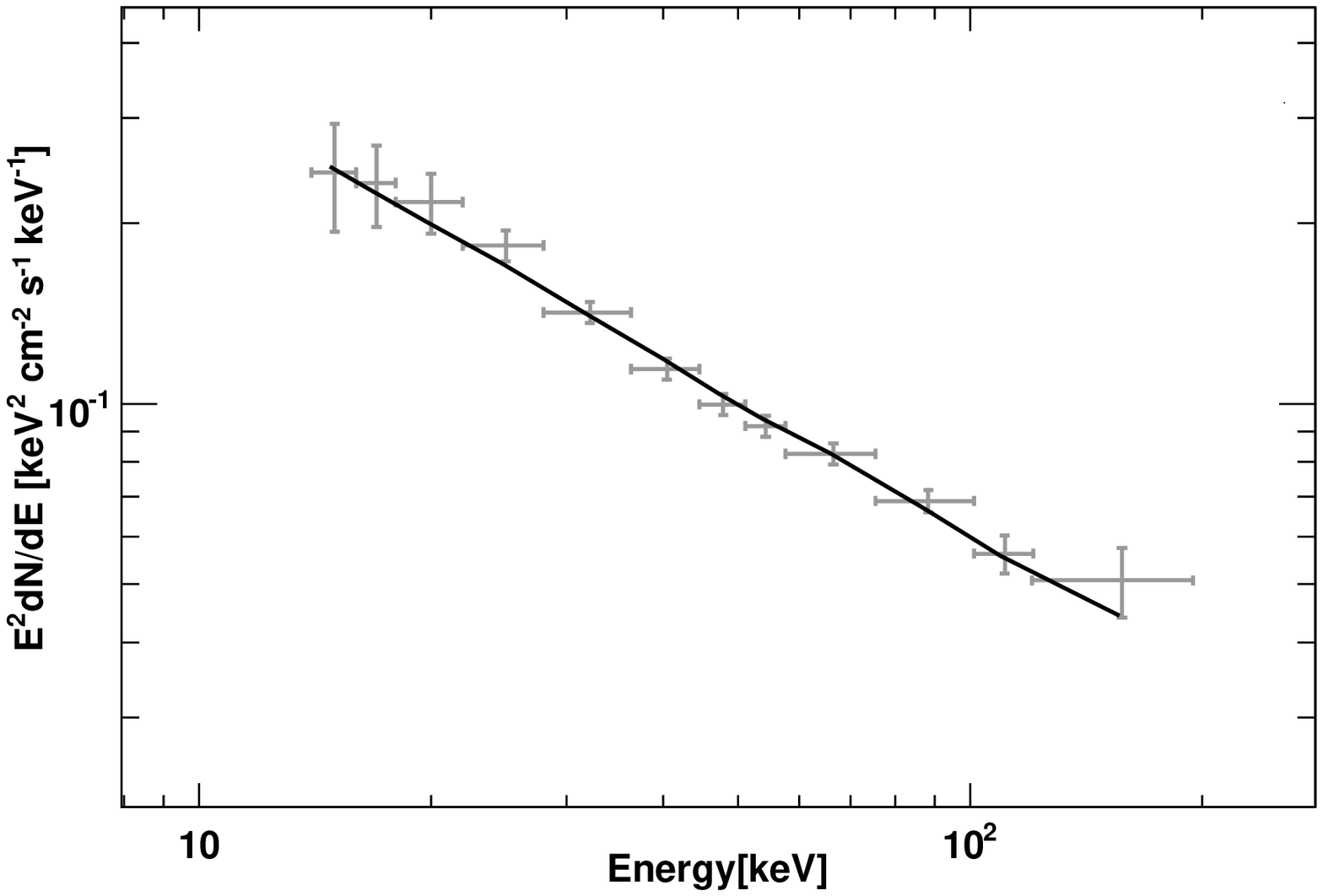} \\
  \end{tabular}
    \end{center}
    \null\vspace{-7mm}
    \caption{Average spectra of 
LMXB with HR$_2$ $>$0.6 (left) and  $<0.6$ (right) respectively 
(see $\S$~\ref{sec:spec} for a definition of the hardness ratio HR$_2$).
The solid lines are the best fit models (power law plus black body for
left and power law for right.)
}
  \label{fig:lmxb}
\end{figure*}

\begin{figure}
\begin{center}
\resizebox{\hsize}{!}{\includegraphics[angle=0]{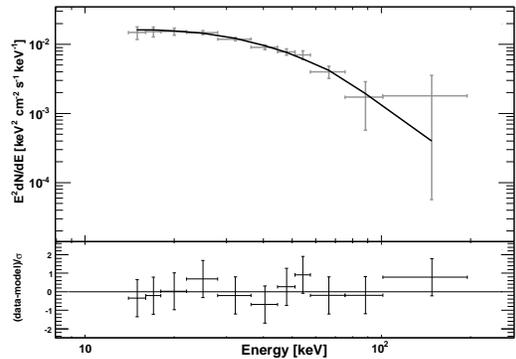}}
\caption{Stacked spectrum of the CVs detected by BAT. The solid line represent 
 the best fit to the data (e.g.  a bremsstrahlung modelwith
 a temperature of 22.68$^{+2.39}_{-2.08}$\,keV.)
}
\label{fig:CV}
\end{center}
\end{figure}

These HRs are defined respectively as:
\begin{equation}
HR_1=\frac{M-H}{M+H}, \ \ \ \ 
HR_2=\frac{S-H} {S+H},
\end{equation}
The hardness ratios, shown in Fig.~\ref{fig:HR}, are normalized to 
the range -1 and +1 setting negative count-rates to zero.\footnote{As shown
in \cite{Ajello09rev}, the analysis of off-source positions showed that, in each energy
channel, the count-rates are found consistent with zero within errors. Thus,
a few negative count-rates consistent with zero (particularly at high energy)
 can be interpreted as a non-detection of the source in that band.
However, as a test we allowed negative count-rates to exist and we found that only
3 objects have an hardness ratio value which falls outside the above range.
All these objects are LMXBs with basically no detection in the hard-band.
Thus, the results reported below do not change whether negative count-rates are set
to zero or not.}
In this plot, hard power-law sources (e.g. AGNs)
occupy the central part of the diagram while soft souces tend to move
to positive values of HR$_1$. 
All the detected sources reported in
Tab.~\ref{tab:cat} are also shown in Fig.~\ref{fig:HR}
(i.e. no sources have been excluded from the graph) 
and this is due to the fact that
BAT is able to constrain efficiently the source spectrum even in the hard band
 (50--195\,keV). 
While it is noticable that all Galactic sources are generally
softer than AGNs, a striking feature is the clustering of
$~20$ LMXBs at large values of both $HR_1$ and $HR_2$.
The combination of the two points to the fact that these sources
exhibit an extremely soft spectrum below 50\,keV and an hard spectrum
above this energy. To investigate this more in details we created
a stacked spectrum of  LMXBs with HR$_2$ $>0.6$ and $<0.6$.
These are shown in Fig.~\ref{fig:lmxb} along with the best-fit models.
Indeed LMXBs clustering in the upper-right corner of the HR plot
exhibit a spectrum which is dominated by a bright
black body component (kT=2.70$\pm0.70$\,keV) at low energy and 
by a flat power law (index of 1.6$\pm0.4$) at high energy.
This corresponds to the high/soft state typical of bright LMXBs.
On the other hand, all the other LMXBs are characterized
by a power-law type spectrum with a photon index of 2.74$\pm0.06$,
corresponding to the low/hard state.
The analysis of the stacked spectra of all Galactic source classes
(with more than 5 objects) is reported in Table~\ref{tab:spec}.
From this analysis it is evident that most Galactic sources
have a non-negligible hard X-ray emission which extends all the way
up to 200\,keV and that can be modeled as a power law. The only
exception is represented by the CV class whose average spectrum is
softer than a power law and  consistent with a bremsstrahlung model with
a temperature of $\sim$22\,keV \citep[see e.g. ][]{brunsch09}.
The stacked spectrum of all the CVs (19) detected by BAT is reported in Fig.~\ref{fig:CV}.

%
%
\section{Variability Analysis}
\label{sec:var}
\begin{figure}
\resizebox{\hsize}{!}{\includegraphics[angle=0]{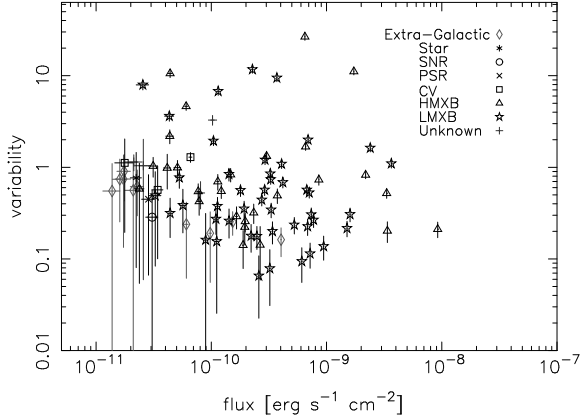}}
\caption{Intrinsic fractional variability ($\sigma_Q/\overline{x}$) of the sources, as a function of the observed count rate, 
for the 7 day time binning. Only objects with a variability greater than 2$\sigma$ are included.
For low count rates even sources with high variability are rejected by this criterion, and for
this reason the lower left corner of the Figure is sparsely populated.
The source types are defined in the catalogue, except for:
Extra-Galactic, which is combined of the Seyfert, Blazar, Galaxy and Galaxy cluster types; Star, which
covers Symbiotic stars, Be stars and a dwarf nova; Unknown, which are all objects not
identified as any of the given types.}
\label{fig:flux}
\end{figure}

\begin{figure}
\resizebox{\hsize}{!}{\includegraphics[angle=0]{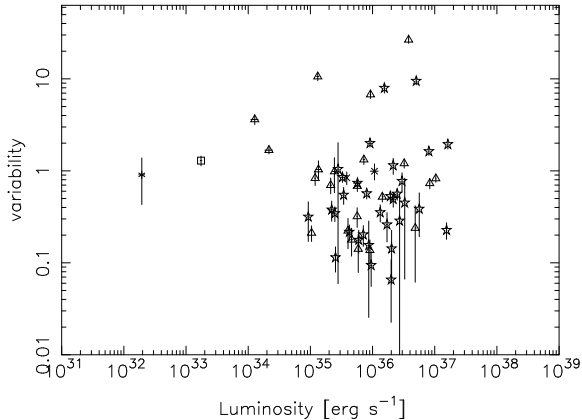}}
\caption{Intrinsic variability ($\sigma_Q$) of the sources, as a function of the observed luminosity,
for the 7 day binning. Only objects with a variability greater than 2$\sigma$ and with 
known distances are included.}
\label{fig:lum}
\end{figure}

To estimate the variability of the sources in our catalogue, we find
numerical maximum-likelihood estimates of the intrinsic variability 
\citep{almaini2000} which has for example been used in the analysis of
AGN from \textit{XMM-Newton} \citep{mateos2007} and \textit{Swift/BAT} 
\citep{beckmann2007} observations. In this method, the intrinsic
variability $\sigma_Q$ is found from solving
\begin{equation}
\displaystyle\sum_{i=0}^N \frac{(x_i-\overline{x})^2-(\sigma_i^2+\sigma_Q^2)}
{(\sigma_i^2+\sigma_Q^2)^2}=0,
\end{equation}
where $x_i$ and $\sigma_i$ are the measured count rate and error in each time bin $i$.
As in \citet{beckmann2007}, we applied this method to the light curves with different
time binnings of 1, 7, 20 and 40 days, and use the variability relative to the average
count rate $\sigma_Q/\overline{x}$ to estimate the strength of the variability. We
simulated random light curves based on the observed source fluxes and errors. These
were used to obtain Monte Carlo estimates of the errors on the calculated variabilities.
We use
both lightcurves generated at random positions and objects that are expected to be
constant (supernova remnants) to investigate systematical effects caused by the instruments
or the data analysis. For the random positions, we find an average variability of
$\sigma_Q$=$6.0\times10^{-5}$, $1.1\times10^{-5}$, $4.5\times10^{-6}$ and $2.5\times10^{-6}$
s$^{-1}$ for 1, 7, 20 and 40 days time binnings, respectively. From the supernova remnants,
the systematic variability is seen to increase with increasing count rate, up to
$\sigma_Q$=$1.4\times10^{-3}$, $9.7\times10^{-4}$, $9.0\times10^{-4}$ and $5.9\times10^{-4}$
s$^{-1}$ for the Crab Nebula. To account for this, we subtracted the variability found
at the random positions from the intrinsic variability of our sources. For the bright
sources with count rates above 10$^{-4}$ s$^{-1}$, we furthermore increased the error on the 
intrinsic variability, by $\sigma_{Crab}\times(\overline{x}-1.0\times10^{-4}$ s$^{-1}$).
In Fig. \ref{fig:flux} we show the intrinsic fractional ($\sigma_Q/\overline{x}$) variability of the sources, as a function
of the observed count rate, for the 7 day time binning. It is clear that almost all of
the strongly variable objects belong to the Galaxy, and the vast majority of these are
either HMXBs or LMXBs. Due to their soft spectra, only few stars
and CVs are detected in the hard X-rays, and
despite extra-Galactic sources being the most numerous source class,
they are almost entirely absent from the plot, as only a very small fraction of them are
variable above the 10\% level \citep{beckmann2007}. Hard
X-ray observations therefore have the potential to classify unidentified X-ray sources.
For the sources with known distances, the luminosities can be
derived. We caution that deriving distances to Galactic sources is
inherently very uncertain and assumption-dependent. Only the bright 
globular cluster sources and a few very well studied binaries and SNR have
distances known to a precision better than 10-20\%, 
whereas the distances to some of the
fainter sources can be uncertain by a factor of $\gtrsim2$. We do not
take the errors into account in our analysis. To significantly impact 
our conclusions, large systematical shifts (factor $\gtrsim5$) of a
high fraction of the sources would be necessary, which is unlikely.
For the sources with calculated luminosities, we plot the variability as a function of source luminosity in Fig. \ref{fig:lum}.
There is no obvious correlation between the luminosity
and the strength of the variability (in the 1-40 days range). Note that the
the sensitivity decreases towards lower fluxes (but depends strongly
on the specific observation pattern). This is the reason that the lower
left parts of Fig. \ref{fig:flux} and \ref{fig:lum} are sparsely populated.

%
%
\section{Source flux distributions}
\label{sec:logn}
\begin{figure}
\resizebox{\hsize}{!}{\includegraphics[angle=0]{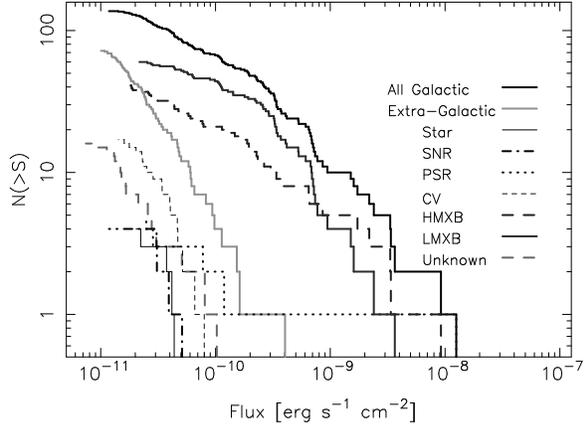}}
\caption{Cumulative $\log$N-$\log$S distributions of the observed sources.}
\label{fig:LNLS}
\end{figure}

\begin{figure}
\resizebox{\hsize}{!}{\includegraphics[angle=0]{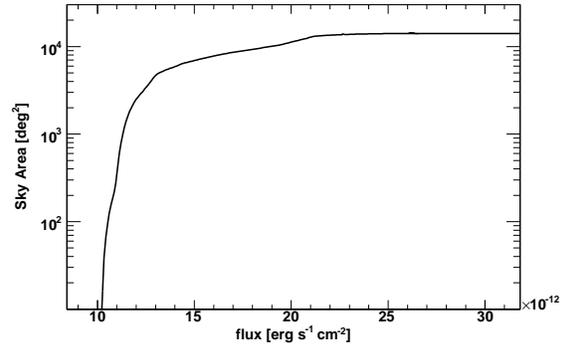}}
\caption{The sky coverage of the survey as a function of flux.
The sensitivity depends on the local exposure and background,
and is worse in the vicinity of bright sources. Above $\sim2\times10^{-11}$
erg s$^{-1}$ cm$^{-2}$ the coverage quicly approaches a value of
$\sim14000$ deg$^{2}$, corresponding to the full area within 20 deg of
latitude from the Galactic plane.}
\label{fig:skycov}
\end{figure}

We use the average fluxes to calculate the source flux distributions
for the different object types. The resulting $\log$N-$\log$S relations
are shown in Figure \ref{fig:LNLS}. The sensitivity of the survey
varies with direction and the sky coverage of our survey 
is shown in Figure
\ref{fig:skycov}. We have not corrected the $\log$N-$\log$S for
the sky coverage, as such a correction depends on the expected
spatial source distribution (see section \ref{sec:lumfun} below for 
LMXBs and HMXBs). The lack of sources at low flux is therefore
obviously caused by the strongly decreasing sky coverage below
$\sim2\times10^{-11}$ erg s$^{-1}$ cm$^{-2}$.
It is clear that at high fluxes
the two most important contributions are the HMXBs and the LMXBs, with
the only exception being the Crab Nebula, which is the object with the
highest flux in our sample. The third most important object type is
the extra-Galactic sources, the contribution of which becomes important
at fluxes below $10^{-10}$ erg s$^{-1}$ cm$^{-2}$. We note that limiting
our survey to the Galactic plane strongly limits the importance of the
extra-Galactic sources. Comparing Figure \ref{fig:LNLS} to Figure
4 of \citet{Grimm2002}, it can be seen that the relative importance
of the HMXBs and the LMXBs is different in our 15-55 keV band to their
results from the 2-10 keV band. In the hard X-rays, the HMXBs dominate
the highest fluxes, with the LMXBs being more important at fluxes
below $10^{-9}$ erg s$^{-1}$, whereas the LMXBs are always dominant in
the soft X-rays. This is due to the fact that the luminous LMXBs have
very soft spectra, and therefore emit almost negligible amounts of
hard X-rays. For example \citet{Revnivtsev2008} found the typical
ratio of hard (17-60 keV) to soft (2-10 keV) luminosities to be
$\sim30$ times lower for LMXBs with luminosities above $2\times10^{37}$
erg s$^{-1}$ than for fainter LMXBs. A similar spectral break is not
seen for the HMXBs.

\section{Luminosity functions of X-ray binaries}
\label{sec:lumfun}
For the majority of bright X-ray binaries in the Galaxy the distances
are known within a factor of 2-3 (see discussion above). 
It is therefore possible to
calculate the luminosity functions of the X-ray binaries. The other
types of objects studied in this paper do not have adequate numbers
of determined distances.\\
The sensitivity of the survey varies with the direction and the luminosity
of the X-ray sources. Following \citet{Grimm2002} we account for this by
setting up a model for the Galaxy and for the range of luminosities
investigated we estimate the fraction of the Galaxy that is visible.
As in \citet{Grimm2002}, we use the three-component model of
\citet{Bahcall1980} for the spatial distribution of the LMXBs, 
consisting of a disk, a bulge and spheroid. The parameters were chosen
to fit the observed distribution of LMXBs \citep[see equations 4--6 and
Table 4 of][]{Grimm2002}, and the disk:bulge:spheroid mass ratios were
chosen to be 2:1:0.8, where the mass of the spheroid is enhanced to account for the
LMXBs formed in globular clusters. As the HMXBs are associated with the
young stellar population in the Galaxy, only the disk component is considered
relevant for the spatial distribution. To account for the spiral structure
of the Galaxy, a spiral model based on optical and radio observations of
giant HII regions \citep{Georgelin1976,Taylor1993} was assumed. This model
consists of 4 spiral arms, which were assumed to have Gaussian density
profiles along the Galactic plane, with a width of 600 pc. The disk
model was modulated by the spiral pattern: 20\% for the LMXBs and
100\% for the HMXBs.\\

For all directions we used the local background to estimate the 
limiting flux detectable by our survey, and used this to create
a sensitivity map. For a given X-ray luminosity and direction, this
enabled us to calculate the maximum distance, for which a X-ray binary
is observable. However, to identify an X-ray source as an X-ray binary, 
and to determine the distance, it is necessary to have an optically 
identified counterpart.
\citet{Grimm2002} estimated that above a distance of 10 kpc from the sun,
the optical identification of X-ray binaries becomes incomplete. We adopt
this result and limit our survey to this distance, irrespective of the
X-ray brightness of the X-ray binaries. However towards the 
galactic bulge, source confusion and extinction are serious and 
optical/IR identifications are incomplete beyond $\sim2-3$ kpc.

\begin{figure}
\resizebox{\hsize}{!}{\includegraphics[angle=0]{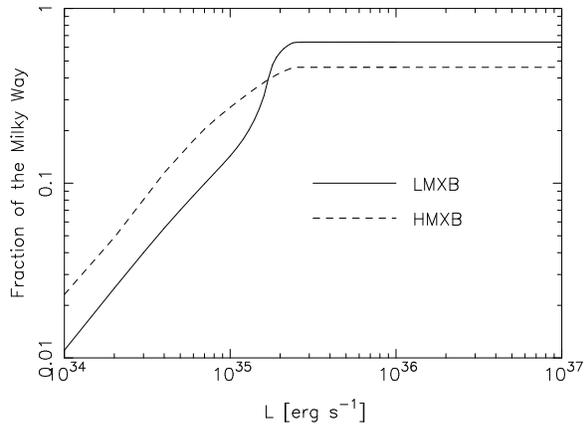}}
\caption{Fraction of the mass of the Galaxy probed by our survey, given the
selection criteria described in the text.}
\label{fig:galaxy}
\end{figure}

\begin{figure}
\resizebox{\hsize}{!}{\includegraphics[angle=0]{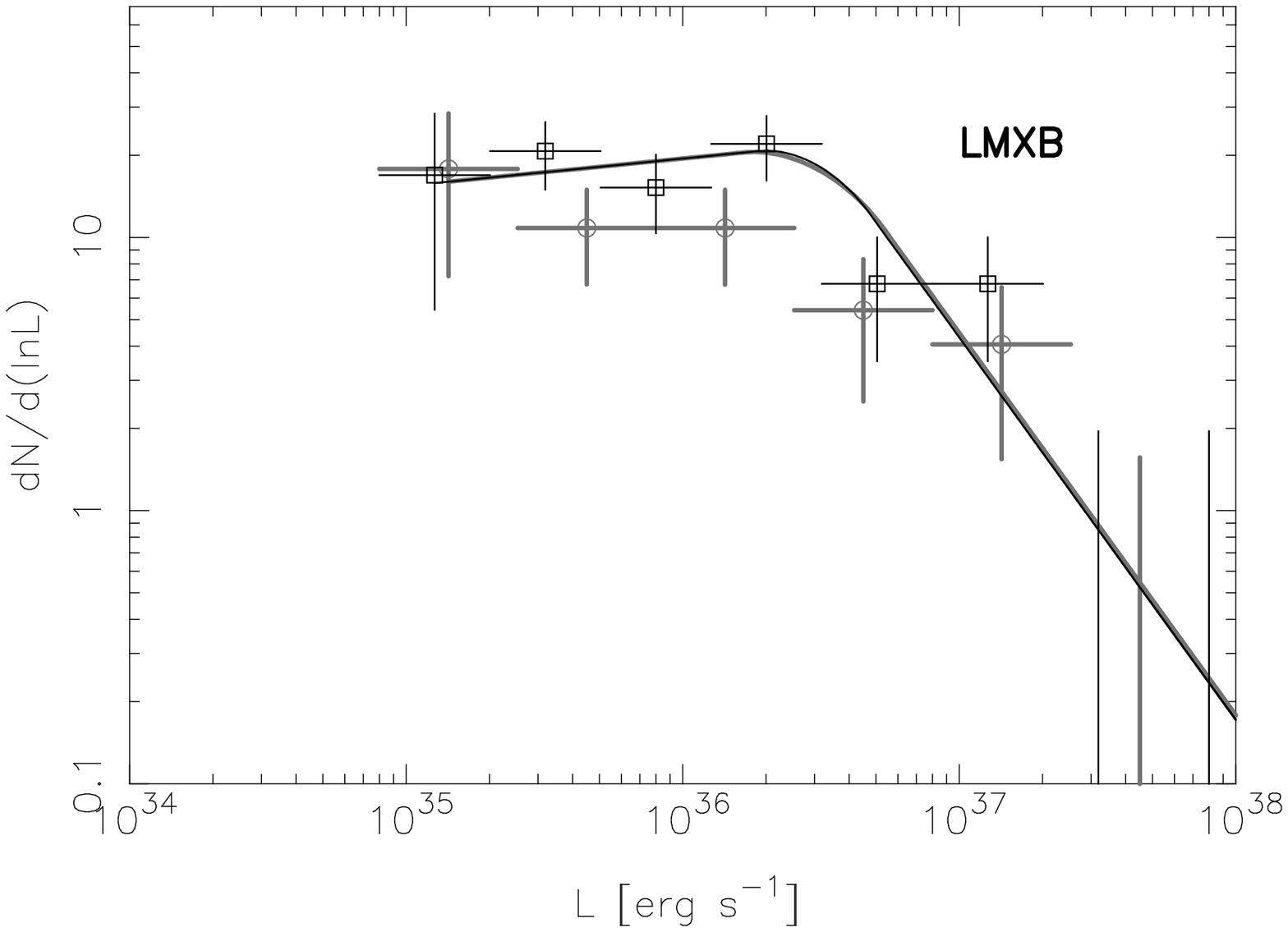}}
\caption{The luminosity function of Galactic LMXBs in the 15-55 keV band, 
corrected for incompleteness, with Poissonian errors. Squares indicate
the results from the entire survey, whereas the inner 10 deg of the bulge
were excluded for the circles. The solid black line 
shows the best maximum likelihood fit to the data from the entire survey, 
see Table \ref{tab:fits}.
}
\label{fig:lumfun_LMXB}
\end{figure}

\begin{figure}
\resizebox{\hsize}{!}{\includegraphics[angle=0]{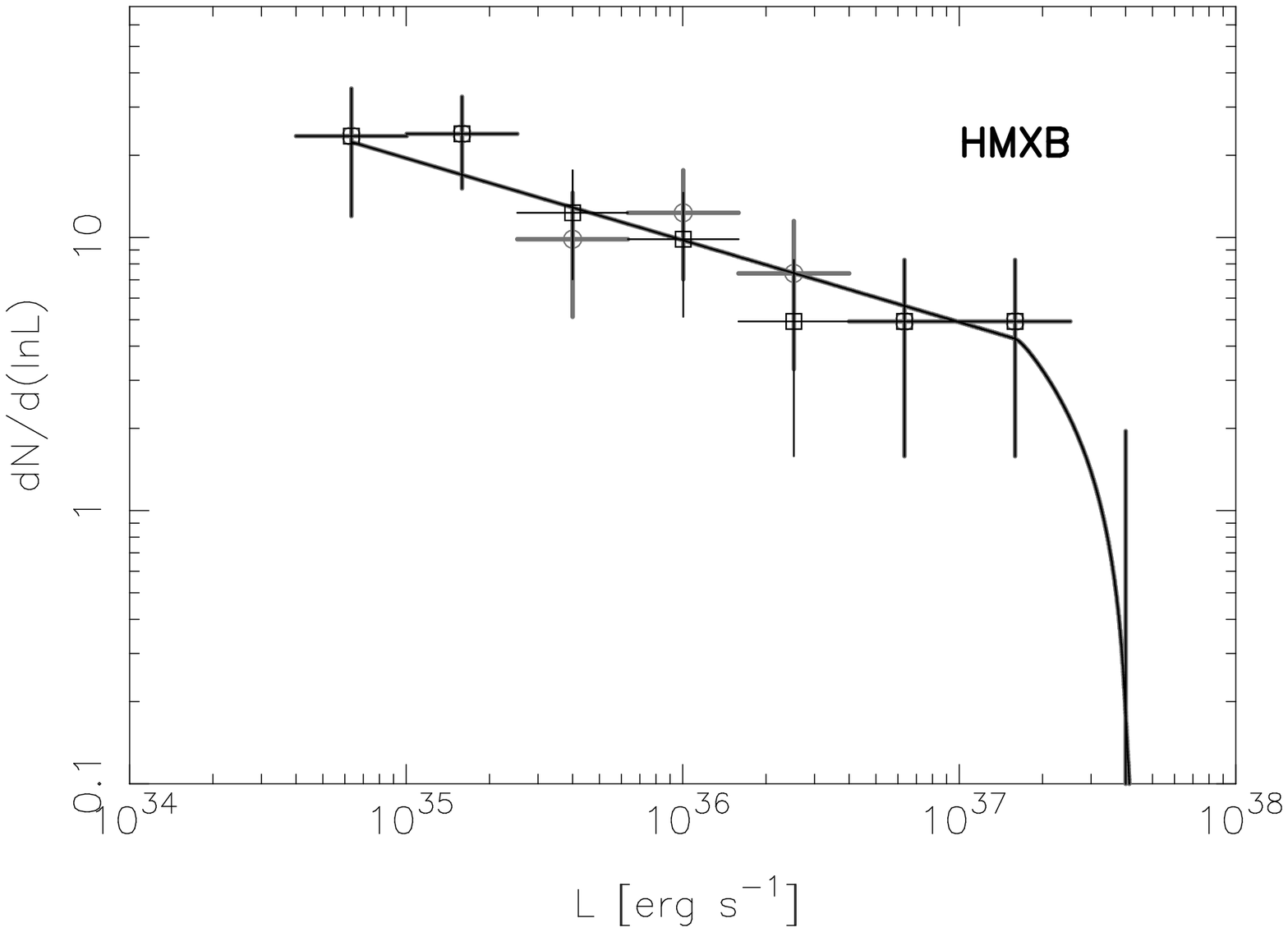}}
\caption{The luminosity function of Galactic HMXBs in the 15-55 keV band,
corrected for incompleteness, with Poissonian errors.
Squares indicate
the results from the entire survey, whereas the inner 10 deg of the bulge
were excluded for the circles. The solid black line 
shows the best maximum likelihood fit to the data from the entire survey, 
see Table
\ref{tab:fits}.}
\label{fig:lumfun_HMXB}
\end{figure}

Combining the X-ray and optical
limits with the model of the Galaxy, we estimate the fraction of the
Galaxy observable as a function of source luminosity. This is shown in
Fig. \ref{fig:galaxy}. Due to our distance constraints and sky coverage,
even the brightest sources are limited to a part of the galaxy, and for
this reason the lines do not reach a value of one.\\
The total Galactic luminosity functions of LMXBs and HMXBs are now found
by correcting the observed luminosity functions for the fraction of the
Galaxy probed by our survey. The outcome is shown for the LMXBs in 
Fig. \ref{fig:lumfun_LMXB} and the HMXBs in Fig. \ref{fig:lumfun_HMXB}.
Also shown in these figures are the luminosity functions obtained
if the inner 10 deg of the bulge are excluded from our analysis, to assess
the effects of source confusion. Obviously the luminosity function of the 
HMXBs is not affected, as these are not concentrated in the bulge. On the
other hand, the luminosity function of the LMXBs is somewhat different with
a lower normalization around $10^{36}$ erg s$^{-1}$. At both lower and higher
luminosities the results are in agreement with the sample including the
inner bulge. This is somewhat surprising as incompleteness due to a lack
of optical IDs is expected to lead to the opposite effect, and could indicate
a higher normalization of LMXBs per unit stellar mass in the bulge than
in the disk. However, the statistical uncertainties,
together with the uncertainties of distance determination and the mass
distribution of the Galaxy (both of which are difficult to quantify), 
are too large for such a conclusion to be significant. We note that
recent results \citep{Kim2010} indicate that the LMXB luminosity functions
are age dependent at bright end ($>10^{38}$ erg s$^{-1}$).

We use Maximum Likelihood (ML) fitting of broken power-laws to analyze 
the shape of the luminosity functions (using the full samples including the
bulge). 
The resulting ranges and slopes are shown in
Table \ref{tab:fits}.
The faint slope of the LMXBs is slightly flatter than, but consistent
with $\Gamma$=1, which is consistent with the INTEGRAL observations of the LMXBs
in the Galactic bulge \citep{Revnivtsev2008}, and with the soft X-ray results
of \citet{Gilfanov2004,Voss-cena,Voss-m31}. Due to the strong spectral
change at luminosities of $\sim10^{37}$ erg s$^{-1}$ 
\citep{Revnivtsev2008}, the LMXB luminosity function breaks and becomes very
steep at higher luminosities. 
The HMXB fit gives a faint slope of $\gamma=-1.3^{+0.3}_{-0.2}$ for the HMXBs. 
This is somewhat shallower, but consistent with the $\sim-1.6$ slope found 
in the soft band both
in the Milky Way \citep{Grimm2002} and in other galaxies \citep{Grimm2003}.
There is a clear break at luminosities above $\sim2\times10^{37}$ 
erg s$^{-1}$, which is
different from the single power-law shape seen in the soft X-rays.
We note that the results are not strongly dependent on the
few brightest sources. If the two brightest LMXB and HMXB sources are 
removed from our samples and the fits are repeated, the best-fit
parameters are within the quoted errors.

\subsection{Total luminosity}
We find the combined luminosity from the Milky Way from summing
all the sources with the individual incompleteness factors. This
gives a hard X-ray luminosity of $1.3\pm0.6\times10^{38}$ erg s$^{-1}$ for
the HMXBs and $1.7\pm0.4\times10^{38}$ erg s$^{-1}$ for the LMXBs.
We note that the total emission is very dependent on the few brightest
X-ray binaries \citep{Gilfanov2004b}.
The total luminosity can also
be found by integrating the broken power-law fits to the observed
sources, which gives $1.4^{+1.5}_{-0.3}\times10^{38}$ erg s$^{-1}$ for
the HMXBs and $1.5^{+1.5}_{-0.3}\times10^{38}$ erg s$^{-1}$ for the
LMXBs. For comparison, the luminosities in the soft band were found
to be $2.0\times10^{38}$ erg s$^{-1}$ for the HMXBs and 
$2.5\times10^{39}$ erg s$^{-1}$ for the LMXBs \citep{Grimm2002}.
As the HMXBs have a relatively hard spectrum, the total luminosity
in the 15-55 keV band is comparable to the soft luminosity, whereas
the soft spectra of the brightest LMXBs causes their total hard band
luminosity to be only 10\% of the soft band luminosity. We
note that, as above, the results from removing the brightest sources
are within the quoted errors.

The number of HMXBs is roughly proportional to the star-formation rate of
a galaxy \citep{Grimm2003}, whereas the LMXBs are related to the stellar
mass of their host galaxy \citep{Gilfanov2004}.
The total luminosities should therefore be compared to the 
star-formation rate in the
Milky Way estimated to be 2-4$M_{\odot}$ yr$^{-1}$ \citep{Diehl2006} and the
stellar mass of $4.8-5.5\times10^{10}$ $M_{\odot}$ \citep{Flynn2006}. 
From this we obtain the ratios: $L_x(HMXB)$/SFR$\sim3-7\times10^{37}$
erg s$^{-1}$ $M_{\odot}^{-1}$ yr and 
$L_x(LMXB)/M_{stellar}\sim3-6\times10^{27}$ erg s$^{-1}$ $M_{\odot}^{-1}$.

A part of the diffuse X-ray background comes from X-ray sources in
galaxies, and our results can be used to calculate the importance.
The local stellar density is $M_{\ast}\sim5\times10^{8}M_{\odot}$Mpc$^{-3}$ 
\citep{Salucci1999,cole2001}, and the local star-formation rate is
$\dot\rho_{\ast}=0.015 M_{\odot}$Mpc$^{-3}$yr$^{-1}$ \citep{Hanish2006}.
This gives a local emissivity from  
X-ray binaries of 
$\sim2-4\times10^{36}$ erg s$^{-1}$ Mpc$^{-3}$, with approximately 80\% coming
from the LMXBs. This can be converted to incident
flux $I_{XRB}$ using equation 19 of \citet{Barcons1995}. Assuming a
normal galaxy density evolution of $(1+z)^3$, we find a flux of 
$1.5-3\times10^{-11}$ erg s$^{-1}$ cm$^{-2}$ Str$^{-1}$.
We therefore conclude
that the contribution to the hard X-ray background, which is 
$9.09\times10^{-8}$ erg cm$^{-2}$ s$^{-1}$ sr$^{-1}$ \citep{ajello08}, 
is negligible. 

It should be noted that because of small-number
statistics, the total X-ray luminosity of the HMXBs in a galaxy
does not scale linearly with the star-formation rate \citep{Grimm2003},
except for galaxies with very high star-formation rates. The Milky Way
might therefore provide a significant under-estimation of the
actual $L_x$/SFR ratio. Indeed \citet{Grimm2002} found a ratio of
$\sim5\times10^{37}$ erg s$^{-1}$ $M_{\odot}^{-1}$ yr, similar to our
results, whereas \citet{Grimm2003} find an average ratio of 
$2\times10^{39}$  erg s$^{-1}$ $M_{\odot}^{-1}$ yr for a large sample
of galaxies\footnote{This value is different from the one listed in
their paper.\citet{Shtykovskiy2004} discussed their star formation
estimates and found that they corresponded to $\sim1/3$ of the total
star formation rate in the $0.1-100 M_{\odot}$ range.}. However, including
this effect still limits the contribution to the X-ray background to
$\lesssim$2\%.

In the soft band galaxies have been found to contribute with $\sim6-12$\% of
the X-ray background \citep[e.g.][]{Ranalli2005}. This is a simple
effect of the fact that the X-ray background is quite hard and so
the X-ray binaries are on average softer. 
Furthermore in the hard band there is no contribution from the 
relatively soft X-rays from diffuse gas
\citep[e.g.][]{Bogdan2008}.

%
%
\section{Discussion}
\label{sec:disc}
Our analysis of the Galaxy in hard X-rays with the \textit{Swift BAT} 
instrument shows that the most important sources are HMXBs and LMXBs,
after which extra-Galactic sources start to dominate the observations.
Compared to the Galaxy in softer X-rays, the contribution to the total
luminosity from HMXBs is much higher, being $\sim$40\% in our observations,
compared to $\lesssim10$\% in the 2-10 keV band \citep{Grimm2002}. This
is mainly due to the fact that luminous LMXBs have very soft spectra,
with only a few percent of the X-rays being hard. Soft X-rays from
HMXBs have been found to be a good indicator of the star-formation rate 
in late-type galaxies \citep{Grimm2003}. Our observations show that hard
X-rays can potentially be more useful for this purpose, 
especially in galaxies with mixed populations, due to the lower
importance of LMXBs. Also in the hard band, possible contributions from
hot X-ray emitting gas are avoided. We note that for nearby
galaxies, where the brightest individual sources might be observed
with future hard X-ray telescopes, the \textit{number} of HMXBs provide
a more reliable estimate of the star formation rate than the total luminosity.
Extra-Galactic observations can be compared to our Milky Way estimates by
integrating the luminosity function given in Table \ref{tab:fits} down
to the observational luminosity limit of the observed galaxy.
At low luminosities, the background AGNs begin to dominate. It is
therefore impossible to infer properties of the populations of weaker 
sources, if optical counterparts have not been observed. 
This will only be possible with instruments with much better spatial
resolution, where individual parts of the Galaxy can be studied in
detail.

%
%
\section{Conclusions}
\label{sec:concl}
We have performed the first survey of the entire Galactic plane
in X-rays, using the \textit{Swift BAT} instrument in the 
15-55 keV energy range. Out of the total 228 sources we identified
the type of 211. The two most important contributions are the 
HMXBs and the LMXBs, both of which
are also among the most variable objects in the Galaxy. The luminosity
function of LMXBs is shown to be consistent with determinations from
soft X-rays, and with previously results from a smaller sample observed
with INTEGRAL. On the other hand, the slope of the luminosity function 
of HMXBs is more shallow than expected. Integrating the total luminosity
of the X-ray binaries and extra-polating to other galaxies, we find that
unresolved populations in galaxies contribute with a relatively small
amount to the hard X-ray background.

\acknowledgments
This research has made use of 
data obtained from the 
High Energy Astrophysics Science Archive Research Center (HEASARC) provided 
by NASA's Goddard Space Flight Center, of the SIMBAD Astronomical Database
which is operated by the Centre de Donn\'ees astronomiques de Strasbourg,
and of the ROSAT All Sky
Survey maintained by the Max Planck Institut f\"ur Extraterrestrische Physik.

{\it Facilities:} \facility{Swift (BAT/XRT)} .

\bibliographystyle{apj}
\bibliography{apj-jour,biblio}


\clearpage
\begin{deluxetable}{lccccclcc}
\tablewidth{0pt}
\tabletypesize{\scriptsize}
\tablecaption{The 228 detected hard X-ray sources \label{tab:cat}}
\tablehead{
\colhead{SWIFT NAME}       & \colhead{R.A.}                &
\colhead{Dec.}            &\colhead{$\sigma$ (pos)}
& \colhead{Flux}                &
\colhead{S/N}              & \colhead{ID}                  &
\colhead{Type}             & \colhead{Offset}              \\
%
%
\colhead{}                        & \colhead{\scriptsize (J2000)}         &
\colhead{\scriptsize (J2000)}     
& \colhead{\scriptsize(arcmin)}
& \colhead{\scriptsize(10$^{-11}$ cgs)} &
\colhead{}                        & \colhead{}                            &
\colhead{}                        & \colhead{\scriptsize(arcmin)}
}
\startdata
J0018.8+8136 & 4.700 & 81.600 & 4.925 & 1.23 & 5.0 & QSO J0017+8135 & BLAZAR & 3.7 \\ 
J0024.9+6407 & 6.235 & 64.128 & 4.341 & 1.33 & 5.5 & 4U 0022+63 & SNR & 2.6 \\ 
J0028.6+5918 & 7.162 & 59.301 & 2.063 & 6.11 & 25.0 & V709 Cas & CV & 1.4 \\ 
J0035.7+5951 & 8.949 & 59.850 & 3.992 & 1.61 & 6.5 & 1ES 0033+59.5 & BLLAC & 1.1 \\ 
J0055.2+4613 & 13.802 & 46.219 & 3.535 & 2.00 & 7.5 & 1RXS J005528.0+461143 & CV & 3.0 \\ 
J0056.5+6042 & 14.127 & 60.705 & 2.544 & 7.11 & 28.6 & Gam Cas & Be star & 1.0 \\ 
J0118.0+6517 & 19.503 & 65.293 & 0.948 & 15.45 & 64.3 & 4U 0114+65 & HMXB & 0.3 \\ 
J0146.5+6144 & 26.635 & 61.745 & 3.158 & 2.49 & 9.7 & PSR J0146+61 & PSR & 1.2 \\ 
J0209.8+5227 & 32.453 & 52.453 & 3.727 & 2.75 & 9.9 & SWIFT J0209.7+5226 & Sy1 & 1.8 \\ 
J0216.2+5126 & 34.051 & 51.449 & 4.090 & 1.69 & 6.0 & SWIFT J0216.3+5128 & Sy2 & 2.8 \\ 
J0217.6+7351 & 34.402 & 73.851 & 4.838 & 1.56 & 6.3 & SWIFT J0218.0+7348 & BLAZAR & 1.5 \\ 
J0240.9+6115 & 40.233 & 61.266 & 4.462 & 1.90 & 7.2 & 2EG J0241+6119 & HMXB & 12.5 \\ 
J0245.0+6228 & 41.266 & 62.473 & 4.233 & 4.67 & 17.9 & QSO B0241+622 & Sy1 & 0.8 \\ 
J0257.3+4535 & 44.350 & 45.600 & 4.973 & 1.49 & 5.1 & \nodata & \nodata & \nodata \\ 
J0319.9+4131 & 49.982 & 41.522 & 2.736 & 6.89 & 23.1 & NGC 1275 & Sy2 & 1.5 \\ 
J0324.9+3409 & 51.248 & 34.152 & 3.607 & 1.73 & 5.9 & 2MASX J03244119+3410459 & Sy1 & 4.1 \\ 
J0325.1+4042 & 51.299 & 40.700 & 5.082 & 1.60 & 5.4 & UGC 2724 & Sy2 & 5.1 \\ 
J0331.2+4354 & 52.801 & 43.901 & 2.796 & 8.90 & 29.9 & GK Per & CV & 0.3 \\ 
J0333.3+3718 & 53.350 & 37.300 & 3.411 & 1.48 & 4.9 & IGR J03334+3718 & Sy1 & 1.1 \\ 
J0334.9+5310 & 53.746 & 53.173 & 0.412 & 83.32 & 314.1 & V0332+53 & HMXB & 0.1 \\ 
J0336.3+3219 & 54.086 & 32.330 & 3.602 & 1.73 & 5.8 & 4C 32.14 & BLAZAR & 2.4 \\ 
J0355.3+3102 & 58.845 & 31.046 & 0.697 & 34.54 & 117.6 & X Per & HMXB & 0.2 \\ 
J0418.4+3801 & 64.605 & 38.029 & 1.772 & 7.44 & 22.7 & 3C 111.0 & Sy1 & 0.8 \\ 
J0443.8+2857 & 70.950 & 28.952 & 5.243 & 2.40 & 6.6 & UGC 3142 & Sy1 & 1.2 \\ 
J0444.1+2813 & 71.033 & 28.221 & 3.353 & 2.56 & 7.0 & SWIFT J0444.1+2813 & Sy2 & 0.3 \\ 
J0452.1+4933 & 73.029 & 49.563 & 2.686 & 3.48 & 11.6 & SWIFT J0452.2+4933 & Sy1 & 1.0 \\ 
J0457.2+4527 & 74.301 & 45.452 & 4.827 & 1.66 & 5.2 & \nodata & \nodata & \nodata \\ 
J0502.7+2445 & 75.699 & 24.753 & 4.029 & 2.03 & 5.4 & Y1062 Tau & CV & 1.8 \\ 
J0502.8+2258 & 75.718 & 22.969 & 4.221 & 2.10 & 5.6 & IRAS 04599+2255 & Sy1 & 2.2 \\ 
J0510.8+1630 & 77.718 & 16.500 & 2.905 & 5.17 & 13.9 & CSV 6150 & Sy1.5 & 1.6 \\ 
J0534.5+2200 & 83.630 & 22.016 & 0.241 & 1267.31 & 3634.0 & Crab & PSR & 0.2 \\ 
J0537.5+2106 & 84.399 & 21.100 & 6.090 & 1.88 & 4.9 & zet Tau & Be* & 3.2 \\ 
J0538.9+2619 & 84.732 & 26.321 & 1.377 & 44.48 & 116.3 & 1A 0535+262 & HMXB & 0.1 \\ 
J0542.6+6050 & 85.673 & 60.842 & 4.427 & 2.31 & 8.7 & BY Cam & CV & 1.4 \\ 
J0552.2-0727 & 88.052 & -7.459 & 1.192 & 16.11 & 43.7 & NGC 2110 & Sy2 & 0.3 \\ 
J0554.9+4626 & 88.740 & 46.442 & 2.778 & 5.39 & 16.2 & MCG +08-11-011 & Sy1.5 & 0.7 \\ 
J0557.8+5353 & 89.471 & 53.896 & 2.819 & 2.28 & 7.9 & V405 Aur & CV & 0.9 \\ 
J0602.2+2828 & 90.553 & 28.482 & 3.448 & 3.80 & 9.0 & SWIFT J0602.2+2829 & Sy1 & 0.9 \\ 
J0602.6+6522 & 90.669 & 65.369 & 3.800 & 1.38 & 5.4 & UGC 3386 & GALAXY & 0.3 \\ 
J0617.1+0908 & 94.291 & 9.138 & 0.600 & 35.35 & 88.7 & 4U 0614+091 & LMXB & 0.6 \\ 
J0623.6-3214 & 95.901 & -32.248 & 6.648 & 1.49 & 5.6 & ESO 426-G 002 & GALAXY-6dF & 2.8 \\ 
J0640.1-2554 & 100.034 & -25.903 & 3.772 & 2.24 & 7.7 & SWIFT J0640.4-2554 & Sy1.2 & 1.0 \\ 
J0655.8+3958 & 103.953 & 39.978 & 4.953 & 2.39 & 6.5 & UGC 3601 & Sy1 & 1.4 \\ 
J0728.8-2606 & 112.224 & -26.101 & 4.787 & 1.67 & 6.1 & 4U 0728-25 & HMXB & 0.4 \\ 
J0731.4+0954 & 112.851 & 9.904 & 3.676 & 2.58 & 6.4 & BG CMi & CV & 2.5 \\ 
J0732.6-1330 & 113.150 & -13.504 & 3.140 & 2.86 & 8.8 & SWIFT J0732.5-1331 & CV & 0.9 \\ 
J0739.3-3143 & 114.849 & -31.725 & 4.351 & 1.53 & 6.0 & LEDA 86063 & Sy2 & 4.4 \\ 
J0746.1-1611 & 116.549 & -16.199 & 5.432 & 2.12 & 7.2 & \nodata & \nodata & \nodata \\ 
J0748.5-6742 & 117.140 & -67.714 & 0.894 & 19.26 & 85.4 & EXO 0748-676 & LMXB & 2.2 \\ 
J0751.0+0324 & 117.750 & 3.400 & 6.050 & 1.80 & 5.0 & LEDA 97223 & Sy1 & 3.3 \\ 
J0751.1+1442 & 117.798 & 14.701 & 2.561 & 3.16 & 8.6 & SWIFT J0750.9+1439 & CV & 2.7 \\ 
J0759.7-3843 & 119.940 & -38.732 & 3.071 & 2.84 & 12.0 & IGR J07597-3842 & Sy1.2 & 0.7 \\ 
J0801.8-4946 & 120.453 & -49.777 & 3.816 & 1.34 & 6.0 & ESO 209-12 & Sy1 & 1.4 \\ 
J0804.1+0505 & 121.031 & 5.094 & 3.259 & 3.17 & 9.3 & UGC 4203 & Sy2 & 1.2 \\ 
J0826.1-7030 & 126.531 & -70.509 & 3.603 & 1.44 & 6.1 & \nodata & \nodata & \nodata \\ 
J0835.2-4511 & 128.808 & -45.190 & 1.806 & 9.17 & 38.6 & Vela Pulsar & PSR & 1.4 \\ 
J0838.3-3559 & 129.597 & -35.998 & 4.216 & 2.09 & 8.8 & Fairall 1146 & Sy1 & 1.5 \\ 
J0839.6-1214 & 129.900 & -12.248 & 3.915 & 1.39 & 5.6 & 3C 206 & Sy1 & 3.6 \\ 
J0902.1-4033 & 135.529 & -40.555 & 0.425 & 344.97 & 1727.0 & Vela X-1 & HMXB & 0.3 \\ 
J0902.5-4810 & 135.648 & -48.177 & 3.912 & 1.36 & 5.5 & IGR J09026-4812 & \nodata & 2.9 \\ 
J0916.1-6218 & 139.038 & -62.308 & 5.416 & 1.84 & 7.4 & SWIFT J0917.2-6221 & Sy1 & 1.0 \\ 
J0920.3-5511 & 140.092 & -55.199 & 2.356 & 5.98 & 23.5 & 4U 0919-54 & LMXB & 0.8 \\ 
J0923.9-3144 & 140.977 & -31.742 & 3.498 & 1.69 & 6.9 & 2MASX J09235371-3141305 & GALAXY6dF & 3.0 \\ 
J0947.5-3056 & 146.890 & -30.947 & 1.746 & 11.34 & 43.1 & ESO 434-40 & Sy2 & 1.4 \\ 
J1009.7-4250 & 152.450 & -42.849 & 6.347 & 1.83 & 6.1 & SWIFT J1009.3-4250 & Sy2 & 2.2 \\ 
J1009.8-5817 & 152.456 & -58.293 & 2.045 & 6.26 & 23.1 & GRO J1008-57 & HMXB & 0.5 \\ 
J1010.9-5749 & 152.739 & -57.827 & 3.248 & 2.03 & 7.3 & SWIFT J1010.1-5747 & STAR/Sym & 1.6 \\ 
J1031.7-3451 & 157.927 & -34.861 & 2.108 & 5.10 & 15.7 & NGC 3281 & Sy2 & 2.0 \\ 
J1038.6-4944 & 159.654 & -49.749 & 4.510 & 1.80 & 5.8 & SWIFT J1038.8-4942 & Sy1 & 2.4 \\ 
J1040.1-4624 & 160.041 & -46.416 & 4.209 & 2.13 & 6.7 & IGR J10404-4625 & Sy2 & 2.2 \\ 
J1121.2-6037 & 170.311 & -60.624 & 0.698 & 89.14 & 397.8 & Cen X-3 & HMXB & 0.2 \\ 
J1131.1-6257 & 172.785 & -62.957 & 2.673 & 4.75 & 15.8 & IGR J11305-6256 & HMXB & 4.2 \\ 
J1143.8-6106 & 175.967 & -61.105 & 2.108 & 3.13 & 10.2 & IGR J11435-6109 & HMXB & 2.7 \\ 
J1147.4-6156 & 176.871 & -61.949 & 1.204 & 24.15 & 80.5 & 1E 1145.1-6141 & HMXB & 0.3 \\ 
J1202.6-5349 & 180.650 & -53.820 & 4.514 & 2.47 & 7.8 & SWIFT J1200.2-5350 & Sy2 & 2.0 \\ 
J1226.6-6246 & 186.658 & -62.768 & 0.300 & 269.38 & 1012.8 & GX 301-2 & LMXB & 0.1 \\ 
J1234.7-6430 & 188.693 & -64.504 & 2.968 & 4.32 & 13.6 & IGR J12349-6434 & STAR/Sym & 3.8 \\ 
J1249.4-5905 & 192.371 & -59.099 & 2.567 & 6.33 & 19.6 & 4U 1246-58 & LMXB & 1.5 \\ 
J1257.4-6915 & 194.365 & -69.264 & 1.841 & 6.32 & 19.3 & 4U 1254-690 & LMXB & 1.7 \\ 
J1300.8-6139 & 195.223 & -61.656 & 3.808 & 1.86 & 5.7 & GX 304-1 & HMXB & 4.3 \\ 
J1301.7-6357 & 195.446 & -63.951 & 2.229 & 4.20 & 12.8 & IGR J13020-6359 & HMXB & 1.7 \\ 
J1305.2-4928 & 196.318 & -49.478 & 1.674 & 10.95 & 32.4 & NGC 4945 & Sy2 & 1.6 \\ 
J1325.4-4300 & 201.363 & -43.016 & 0.643 & 40.32 & 116.2 & Cen A & Sy2 & 0.2 \\ 
J1326.2-6207 & 201.575 & -62.127 & 1.521 & 14.26 & 41.9 & 4U 1323-619 & LMXB & 2.2 \\ 
J1347.5-6033 & 206.884 & -60.566 & 2.193 & 4.76 & 13.5 & 4U 1344-60 & Sy1 & 3.1 \\ 
J1413.3-6518 & 213.337 & -65.314 & 1.356 & 16.24 & 44.3 & Cir Galaxy & Sy2 & 2.0 \\ 
J1451.6-5537 & 222.916 & -55.632 & 5.584 & 1.97 & 5.0 & IGR J14515-5542 & Sy2 & 2.9 \\ 
J1453.6-5523 & 223.402 & -55.400 & 6.204 & 2.01 & 5.0 & IGR J14536-5522 & CV & 2.4 \\ 
J1457.9-4305 & 224.500 & -43.100 & 4.518 & 2.23 & 5.0 & IC 4518 & Sy2 & 3.3 \\ 
J1514.1-5908 & 228.549 & -59.145 & 1.169 & 11.89 & 29.4 & PSR B1509-58 & PSR & 2.2 \\ 
J1520.6-5710 & 230.153 & -57.175 & 1.187 & 15.58 & 38.5 & Cir X-1 & LMXB & 0.7 \\ 
J1542.3-5222 & 235.592 & -52.383 & 0.719 & 32.11 & 79.0 & 4U 1538-522 & HMXB & 0.3 \\ 
J1547.9-6233 & 237.000 & -62.554 & 2.861 & 5.87 & 14.2 & 4U 1543-62 & LMXB & 1.2 \\ 
J1548.2-4529 & 237.053 & -45.484 & 2.186 & 6.27 & 14.1 & IGR J15479-4529 & CV & 0.5 \\ 
J1601.1-6045 & 240.300 & -60.766 & 5.361 & 2.29 & 5.5 & 1H 1556-605 & LMXB & 2.0 \\ 
J1612.1-6034 & 243.049 & -60.578 & 4.779 & 2.45 & 5.8 & IGR J16119-6036 & Sy1 & 4.1 \\ 
J1612.7-5225 & 243.188 & -52.422 & 1.190 & 40.98 & 111.8 & 4U 1608-522 & LMXB & 0.3 \\ 
J1619.3-4944 & 244.829 & -49.738 & 5.762 & 2.59 & 6.0 & IGR J16195-4945 & HMXB & 2.2 \\ 
J1619.6-2807 & 244.905 & -28.124 & 3.311 & 4.17 & 8.4 & IGR J16194-2810 & STAR/Sym & 0.9 \\ 
J1620.8-5130 & 245.215 & -51.508 & 3.779 & 2.95 & 6.8 & IGR J16207-5129 & HMXB & 2.3 \\ 
J1626.5-5156 & 246.645 & -51.936 & 4.005 & 4.32 & 10.0 & SWIFT J1626.6-5156 & PSR & 0.5 \\ 
J1628.1-4912 & 247.027 & -49.211 & 1.871 & 9.87 & 22.7 & 4U 1624-490 & LMXB & 1.0 \\ 
J1631.8-4848 & 247.952 & -48.816 & 1.027 & 37.44 & 91.3 & IGR J16318-4848 & HMXB & 0.8 \\ 
J1632.0-4752 & 248.015 & -47.875 & 1.282 & 22.59 & 53.1 & AX J1631.9-4752 & PSR & 0.3 \\ 
J1632.3-6725 & 248.097 & -67.422 & 1.165 & 26.13 & 62.3 & 4U 1626-67 & LMXB & 2.5 \\ 
J1638.8-6423 & 249.714 & -64.400 & 4.227 & 2.67 & 6.3 & CIZA J1638.2-6420 & GCluster & 4.9 \\ 
J1639.1-4642 & 249.798 & -46.702 & 2.315 & 7.42 & 17.0 & IGR J16393-4643 & HMXB & 2.2 \\ 
J1640.9-5345 & 250.229 & -53.755 & 0.752 & 42.21 & 106.1 & 4U 1636-536 & LMXB & 0.2 \\ 
J1642.0-4532 & 250.506 & -45.538 & 2.797 & 5.09 & 11.7 & IGR J16418-4532 & \nodata & 3.1 \\ 
J1645.8-4536 & 251.451 & -45.610 & 0.977 & 77.15 & 241.1 & GX 340+0 & LMXB & 0.1 \\ 
J1648.4-3035 & 252.108 & -30.590 & 2.270 & 3.86 & 8.5 & 2MASX J16481523-3035037 & Sy1 & 2.3 \\ 
J1649.5-4350 & 252.395 & -43.849 & 3.474 & 3.05 & 7.0 & IGR J16493-4348 & XBinary & 3.4 \\ 
J1651.8-5914 & 252.968 & -59.239 & 4.347 & 2.29 & 5.3 & ESO 138-1 & Sy2 & 4.1 \\ 
J1654.0-3950 & 253.500 & -39.846 & 0.965 & 22.75 & 54.6 & GRO J1655-40 & LMXB & 0.0 \\ 
J1654.8-1920 & 253.701 & -19.348 & 4.775 & 2.57 & 5.5 & 1RXS J165443.5-191620 & \nodata & 4.7 \\ 
J1656.2-3301 & 254.055 & -33.031 & 3.544 & 3.01 & 6.8 & SWIFT J1656.3-3302 & BLAZAR & 0.8 \\ 
J1700.2-4221 & 255.056 & -42.351 & 4.246 & 2.80 & 6.5 & AX J1700.2-4220 & HMXB & 1.1 \\ 
J1700.8-4139 & 255.203 & -41.658 & 0.782 & 68.21 & 196.6 & OAO 1657-415 & HMXB & 0.9 \\ 
J1700.9-4611 & 255.246 & -46.184 & 1.516 & 35.96 & 88.7 & XTE J1701-462 & LMXB & 0.1 \\ 
J1702.8-4847 & 255.707 & -48.789 & 0.940 & 58.45 & 145.7 & GX 339-4 & LMXB & 0.0 \\ 
J1703.9-3750 & 255.987 & -37.843 & 0.425 & 359.13 & 792.6 & 4U 1700-377 & HMXB & 0.2 \\ 
J1705.7-3625 & 256.437 & -36.423 & 0.568 & 94.45 & 328.3 & GX 349+2 & LMXB & 0.1 \\ 
J1706.3-6144 & 256.578 & -61.745 & 3.843 & 2.78 & 6.4 & IGR J17062-6143 & TRANSIENT & 2.7 \\ 
J1706.2-4302 & 256.562 & -43.040 & 1.134 & 29.03 & 71.0 & 4U 1702-429 & LMXB & 0.2 \\ 
J1708.8-3218 & 257.202 & -32.303 & 5.347 & 2.83 & 6.6 & 4U 1705-32 & LMXB & 1.5 \\ 
J1708.8-4406 & 257.213 & -44.113 & 1.170 & 32.79 & 80.3 & 4U 1705-440 & LMXB & 0.9 \\ 
J1709.8-3626 & 257.450 & -36.449 & 5.029 & 2.12 & 5.0 & IGR J17091-3624 & XRB & 8.8 \\ 
J1710.2-2806 & 257.566 & -28.111 & 3.773 & 3.26 & 7.6 & XTE J1710-281 & LMXB & 1.5 \\ 
J1712.3-2319 & 258.091 & -23.324 & 2.895 & 8.51 & 19.7 & Oph Cluster & GCluster & 1.7 \\ 
J1712.5-4050 & 258.146 & -40.847 & 3.258 & 2.73 & 6.5 & 4U 1708-40 & LMXB & 2.3 \\ 
J1712.6-3736 & 258.163 & -37.611 & 4.146 & 4.44 & 10.6 & SAX J1712.6-3739 & LMXB & 2.2 \\ 
J1712.7-2417 & 258.181 & -24.285 & 3.080 & 4.31 & 10.0 & V2400 Oph & CV & 2.9 \\ 
J1717.0-6249 & 259.258 & -62.833 & 2.148 & 6.17 & 14.3 & NGC 6300 & Sy2 & 0.9 \\ 
J1719.7-4100 & 259.933 & -41.009 & 2.882 & 3.42 & 8.2 & 1RXS J171935.6-410054 & CV & 1.6 \\ 
J1720.2-3112 & 260.063 & -31.214 & 4.073 & 2.75 & 6.6 & IGR J17200-3116 & HMXB & 4.6 \\ 
J1725.2-3617 & 261.303 & -36.289 & 3.448 & 9.29 & 22.4 & IGR J17252-3616 & HMXB & 0.5 \\ 
J1727.5-3048 & 261.899 & -30.806 & 2.335 & 22.15 & 53.6 & 4U 1722-30 & LMXB & 0.6 \\ 
J1730.3-0559 & 262.599 & -5.998 & 3.086 & 5.11 & 11.6 & IGR J17303-0601 & CV & 0.6 \\ 
J1731.7-1657 & 262.938 & -16.950 & 0.826 & 37.29 & 90.2 & GX 9+9 & LMXB & 0.7 \\ 
J1731.9-3349 & 262.995 & -33.829 & 0.626 & 67.38 & 195.5 & 4U 1728-34 & LMXB & 0.5 \\ 
J1732.0-2444 & 263.016 & -24.744 & 0.637 & 70.82 & 218.7 & GX 1+4 & LMXB & 0.4 \\ 
J1737.4-2908 & 264.375 & -29.147 & 2.771 & 5.82 & 14.3 & GRS 1734-292 & Sy1 & 2.3 \\ 
J1738.3-2659 & 264.578 & -26.988 & 1.048 & 13.57 & 33.3 & SLX 1735-269 & LMXB & 1.2 \\ 
J1738.9-4427 & 264.749 & -44.452 & 1.157 & 61.11 & 170.1 & 4U 1735-44 & LMXB & 0.3 \\ 
J1740.4-2821 & 265.109 & -28.357 & 5.578 & 6.48 & 15.9 & SLX 1737-282 & LMXB & 7.3 \\ 
J1743.9-2944 & 265.981 & -29.738 & 1.759 & 32.53 & 84.0 & 1E 1740.7-2943 & LMXB & 0.5 \\ 
J1745.9-2930 & 266.489 & -29.500 & 2.083 & 19.27 & 46.4 & 2E 1742.9-2929 & LMXB & 2.0 \\ 
J1746.2-3213 & 266.556 & -32.225 & 4.290 & 3.06 & 7.6 & IGR J17464-3213 & LMXB & 15.1 \\ 
J1746.3-2851 & 266.598 & -28.863 & 1.512 & 12.02 & 29.6 & 1E 1742.9-2849 & LMXB & 3.1 \\ 
J1747.2-3001 & 266.811 & -30.029 & 2.154 & 11.15 & 26.2 & AX J1747.4-3000 & LMXB & 3.0 \\ 
J1747.9-2633 & 266.993 & -26.563 & 0.997 & 26.66 & 65.9 & GX 3+1 & LMXB & 0.5 \\ 
J1748.9-3257 & 267.230 & -32.959 & 5.932 & 2.34 & 5.7 & IGR J17488-3253 & Sy1 & 2.8 \\ 
J1749.6-2820 & 267.410 & -28.348 & 2.433 & 10.25 & 26.0 & IGR J17497-2821 & \nodata & 0.4 \\ 
J1750.2-3704 & 267.557 & -37.072 & 2.833 & 6.54 & 15.9 & 4U 1746-37 & LMXB & 1.2 \\ 
J1753.4-0126 & 268.373 & -1.447 & 0.536 & 69.58 & 188.3 & SWIFT J1753.5-0127 & LMXB & 0.5 \\ 
J1759.8-2200 & 269.964 & -22.007 & 2.538 & 5.28 & 13.1 & XTE J1759-220 & LMXB & 1.3 \\ 
J1801.1-2504 & 270.285 & -25.079 & 0.956 & 193.47 & 475.2 & GX 5-1 & LMXB & 0.1 \\ 
J1801.2-2544 & 270.304 & -25.740 & 0.810 & 68.51 & 213.6 & GRS 1758-258 & LMXB & 0.2 \\ 
J1801.5-2031 & 270.393 & -20.522 & 0.536 & 53.03 & 140.9 & GX 9+1 & LMXB & 0.6 \\ 
J1808.6-2024 & 272.165 & -20.417 & 5.660 & 2.85 & 7.1 & SGR 1806-20 & PSR & 0.3 \\ 
J1814.5-1708 & 273.628 & -17.148 & 0.731 & 35.37 & 88.1 & GX 13+1 & LMXB & 0.6 \\ 
J1815.1-1205 & 273.784 & -12.093 & 1.517 & 32.64 & 79.1 & 4U 1812-12 & LMXB & 1.1 \\ 
J1816.0-1402 & 274.013 & -14.037 & 0.278 & 199.08 & 546.7 & GX 17+2 & LMXB & 0.4 \\ 
J1817.6-3300 & 274.422 & -33.011 & 1.614 & 10.44 & 25.4 & XTE J1817-330 & LMXB & 0.6 \\ 
J1821.4-1317 & 275.350 & -13.300 & 4.370 & 2.06 & 5.0 & IGR J18214-1318 & \nodata & 0.7 \\ 
J1823.6-3021 & 275.925 & -30.360 & 1.093 & 76.69 & 250.2 & 4U 1820-30 & LMXB & 0.3 \\ 
J1825.3-0001 & 276.345 & -0.026 & 2.916 & 5.67 & 14.5 & 4U 1822-000 & LMXB & 0.8 \\ 
J1825.8-3706 & 276.455 & -37.106 & 0.858 & 44.39 & 107.6 & 4U 1822-371 & LMXB & 0.5 \\ 
J1829.5-2347 & 277.378 & -23.798 & 1.284 & 85.25 & 313.8 & Ginga 1826-24 & LMXB & 0.6 \\ 
J1833.5-1034 & 278.378 & -10.582 & 2.643 & 4.36 & 10.5 & SNR 021.5-00.9 & SNR & 1.4 \\ 
J1833.7-2103 & 278.449 & -21.053 & 4.978 & 2.76 & 6.6 & PKS 1830-21 & BLAZAR & 1.9 \\ 
J1834.9+3240 & 278.745 & 32.677 & 2.435 & 4.64 & 17.4 & 3C 382 & Sy1 & 1.5 \\ 
J1835.7-3259 & 278.938 & -32.992 & 1.429 & 13.24 & 31.0 & XB 1832-330 & LMXB & 0.2 \\ 
J1837.9-0654 & 279.496 & -6.916 & 4.031 & 3.09 & 7.5 & HESS J1837-069 & SNR & 4.1 \\ 
J1839.9+0502 & 279.997 & 5.039 & 0.586 & 34.47 & 99.7 & Ser X-1 & LMXB & 0.5 \\ 
J1841.4-0457 & 280.351 & -4.950 & 3.486 & 3.07 & 7.6 & SNR 027.4+00.0 & SNR & 1.5 \\ 
J1845.6+0051 & 281.422 & 0.855 & 3.408 & 4.37 & 11.6 & Ginga 1843+009 & HMXB & 2.3 \\ 
J1846.4-0302 & 281.602 & -3.046 & 3.900 & 2.49 & 6.3 & PSR J1846-0258 & PSR & 4.3 \\ 
J1848.2-0310 & 282.074 & -3.174 & 3.532 & 5.32 & 13.4 & IGR J18483-0311 & PSR & 0.7 \\ 
J1853.0-0842 & 283.268 & -8.703 & 1.962 & 8.94 & 21.3 & 4U 1850-086 & LMXB & 0.2 \\ 
J1854.9-3109 & 283.733 & -31.164 & 3.499 & 8.60 & 19.0 & V1223 Sgr & CV & 1.4 \\ 
J1855.4-0236 & 283.875 & -2.608 & 1.007 & 16.55 & 42.4 & XTE J1855-026 & HMXB & 0.3 \\ 
J1856.1+1536 & 284.050 & 15.600 & 5.688 & 1.39 & 4.8 & 2E 1853.7+1534 & Sy1 & 3.6 \\ 
J1858.8+0322 & 284.716 & 3.369 & 5.875 & 1.87 & 5.3 & XTE J1858+034 & HMXB & 4.2 \\ 
J1900.1-2453 & 285.029 & -24.900 & 0.900 & 38.78 & 84.5 & HETE J1900.1-2455 & LMXB & 1.3 \\ 
J1901.6+0127 & 285.410 & 1.459 & 4.193 & 3.88 & 10.6 & XTE J1901+014 & HMXB & 1.4 \\ 
J1909.6+0950 & 287.415 & 9.837 & 1.641 & 21.74 & 71.8 & 4U 1907+09 & HMXB & 0.6 \\ 
J1910.7+0735 & 287.693 & 7.594 & 1.916 & 19.64 & 61.2 & 4U 1909+07 & HMXB & 0.5 \\ 
J1911.2+0034 & 287.816 & 0.581 & 2.092 & 11.48 & 32.2 & Aql X-1 & LMXB & 0.2 \\ 
J1911.7+0500 & 287.940 & 5.000 & 2.884 & 10.17 & 30.0 & SS 433 & HMXB & 1.4 \\ 
J1914.0+0952 & 288.525 & 9.883 & 1.428 & 11.40 & 37.4 & IGR J19140+0951 & HMXB & 1.0 \\ 
J1915.1+1056 & 288.795 & 10.945 & 1.030 & 380.66 & 1363.0 & GRS 1915+105 & LMXB & 0.2 \\ 
J1918.7-0514 & 289.690 & -5.243 & 1.758 & 12.82 & 31.5 & 4U 1916-053 & LMXB & 0.6 \\ 
J1920.9+4357 & 290.240 & 43.965 & 3.325 & 2.87 & 10.7 & ABELL 2319 & GCluster & 2.2 \\ 
J1922.5-1715 & 290.640 & -17.266 & 2.235 & 9.61 & 20.0 & SWIFT J1922.7-1716 & XBinary & 1.3 \\ 
J1924.5+5016 & 291.131 & 50.268 & 3.487 & 2.23 & 8.3 & CH Cyg & STAR/Sym & 1.0 \\ 
J1930.2+3411 & 292.568 & 34.184 & 2.895 & 1.75 & 6.8 & SWIFT J1930.5+3414 & Sy1 & 0.6 \\ 
J1940.2-1024 & 295.052 & -10.410 & 3.946 & 4.65 & 10.6 & V1432 Aql & CV & 0.9 \\ 
J1942.5-1018 & 295.650 & -10.306 & 3.809 & 4.77 & 10.9 & NGC 6814 & Sy1 & 1.5 \\ 
J1943.9+2119 & 295.991 & 21.322 & 3.347 & 1.74 & 6.5 & RX J1943.9+2118 & \nodata & 1.8 \\ 
J1949.5+3012 & 297.393 & 30.206 & 2.600 & 8.51 & 34.4 & KS 1947+300 & HMXB & 0.8 \\ 
J1955.6+3205 & 298.922 & 32.099 & 1.249 & 30.19 & 128.3 & 4U 1954+31 & HMXB & 0.3 \\ 
J1958.3+3512 & 299.592 & 35.200 & 0.333 & 930.01 & 4123.6 & Cyg X-1 & HMXB & 0.3 \\ 
J1959.4+4044 & 299.865 & 40.735 & 2.369 & 6.48 & 24.4 & Cyg A & Sy2 & 0.2 \\ 
J1959.5+6509 & 299.889 & 65.159 & 2.973 & 2.53 & 10.2 & 2MASX J19595975+6508547 & BLAZAR & 2.9 \\ 
J2000.3+3211 & 300.088 & 32.186 & 2.896 & 2.37 & 9.4 & SWIFT J2000.6+3210 & Be Star/HXB & 0.3 \\ 
J2018.5+4042 & 304.650 & 40.700 & 4.246 & 1.45 & 5.4 & IGR J20187+4041 & AGN & 1.1 \\ 
J2028.4+2544 & 307.121 & 25.743 & 3.447 & 3.47 & 13.1 & MCG+04-48-002 & Sy2 & 1.5 \\ 
J2032.2+3738 & 308.065 & 37.638 & 0.321 & 196.91 & 573.9 & EXO 2030+375 & HMXB & 0.1 \\ 
J2032.4+4057 & 308.112 & 40.952 & 0.284 & 237.59 & 1012.0 & Cyg X-3 & HMXB & 0.4 \\ 
J2033.4+2144 & 308.354 & 21.749 & 3.788 & 1.67 & 6.2 & XSS J20348+2157 & GALAXY & 2.2 \\ 
J2036.9+4150 & 309.248 & 41.849 & 4.746 & 1.55 & 5.8 & SWIFT J2037.2+4151 & \nodata & 2.3 \\ 
J2042.7+7508 & 310.678 & 75.137 & 2.577 & 3.28 & 13.5 & 4C +74.26 & BLAZAR & 0.4 \\ 
J2056.6+4942 & 314.153 & 49.700 & 5.060 & 1.32 & 5.2 & \nodata & \nodata & \nodata \\ 
J2123.4+4216 & 320.853 & 42.277 & 5.518 & 1.43 & 5.6 & V2069 Cyg & CV & 4.0 \\ 
J2124.5+5058 & 321.134 & 50.979 & 2.055 & 9.37 & 38.6 & IGR J21247+5058 & Sy1 & 1.2 \\ 
J2127.4+5656 & 321.862 & 56.938 & 2.096 & 2.72 & 11.7 & SWIFT J2127.4+5654 & Sy1 & 2.5 \\ 
J2133.6+5107 & 323.401 & 51.124 & 2.806 & 3.97 & 16.5 & RX J2133.7+5107 & CV & 1.2 \\ 
J2136.0+4730 & 324.018 & 47.512 & 4.468 & 1.44 & 5.9 & RX J2135.9+4728 & Sy1 & 2.8 \\ 
J2138.5+3208 & 324.634 & 32.138 & 3.314 & 1.45 & 5.7 & LEDA 67084 & Sy1 & 3.2 \\ 
J2142.6+4334 & 325.659 & 43.574 & 2.336 & 4.36 & 17.7 & 3A 2140+433 & DWARF NOVA & 2.9 \\ 
J2144.6+3819 & 326.171 & 38.322 & 0.678 & 60.01 & 322.0 & Cyg X-2 & LMXB & 0.0 \\ 
J2202.8+4216 & 330.701 & 42.271 & 4.589 & 1.57 & 6.5 & QSO B2200+420 & BLLAC & 1.0 \\ 
J2208.0+5430 & 332.004 & 54.511 & 0.877 & 15.43 & 68.7 & 4U 2206+54 & HMXB & 0.8 \\ 
J2245.5+3941 & 341.388 & 39.687 & 3.597 & 1.75 & 7.6 & 3C 452 & Sy2 & 3.1 \\ 
J2258.9+4051 & 344.730 & 40.863 & 3.257 & 1.36 & 5.9 & UGC 12282 & Sy1 & 4.2 \\ 
J2307.8+4012 & 346.950 & 40.200 & 6.817 & 1.15 & 5.0 & 1RXS J230757.5+401636 & \nodata & 5.0 \\ 
J2323.2+5848 & 350.819 & 58.811 & 2.034 & 6.25 & 26.9 & Cas A & SNR & 1.0 \\
\enddata
\end{deluxetable}

\begin{deluxetable}{lccclcc}
\tablewidth{0pt}
\tablecaption{Tentative ID for Unidentified Sources \label{tab:uid}}
\tablehead{
\colhead{SWIFT NAME}       & \colhead{R.A.}                &
\colhead{Decl.}            & \colhead{S/N}                 & 
\colhead{ID}               &\colhead{Type}                 & 
\colhead{Offset}              \\
%
%
\colhead{}                        & \colhead{\scriptsize (J2000)}     &
\colhead{\scriptsize (J2000)}     & \colhead{}                        & 
\colhead{}                        & \colhead{}                        & 
\colhead{\scriptsize(arcmin)}
}
\startdata

J0457.2+4527 & 74.301 & 45.452   & 5.2   & 1RXS J045707.4+452751 & AGN\tablenotemark{a} & 1.1\\
J0746.1-1611 & 116.549 & -16.199 & 7.2   &1RXS J074616.8-161127 & & 1.3 \\
J0826.1-7030 & 126.531 & -70.509 & 6.1   &1ES 0826-70.3 & &1.7 \\
J2056.6+4942 & 314.153 & 49.700  & 5.2   & RX J2056.6+4940& AGN\tablenotemark{b} & 2.0\\
\enddata

\tablenotetext{a}{The extragalactic nature of 1RXS J045707.4+452751 has
been proposed by \cite{kaplan06} on the basis of the hard X-ray spectrum and 
the X-ray-to-IR flux ratio.}
\tablenotetext{b}{The nature of RX J2056.6+4940 is likely extragalactic
because of its association with  a radio-loud object \citep{brinkmann97}.}

\end{deluxetable}

\begin{deluxetable}{lc}
\tablewidth{0pt}
\tablecaption{Numbers of different source types in our catalogue.}
\tablehead{
\colhead{CLASS} & \colhead{Number}}
\startdata
LMXB & 61\\
HMXB & 43\\
CV & 19\\
Supernova remnant & 6\\
Pulsar & 6\\
Star: symbiotic & 4\\
Star: Dwarf Nova & 1\\
AGN: Seyfert & 56\\
AGN: BL Lac & 2\\
AGN: Blazar & 7\\
AGN: undefined & 5\\
Galaxy cluster & 3\\
\enddata
\label{tab:objtyp}
\end{deluxetable}

\begin{deluxetable}{lccc}
\tablewidth{0pt}
\tablecaption{Average spectral properties of Galactic sources\label{tab:spec}
derived from the analysis of stacked spectra.}
\tablehead{
\colhead{CLASS} & \colhead{N. Objects} & \colhead{Photon index} 
& \colhead{kT\tablenotemark{a} (keV)} }
\startdata
LMXB$_({HR_2>0.6})$ & 21 & 2.25$^{+0.83}_{-0.82}$ & 2.7$^{+0.19}_{-0.25}$\\
LMXB$_({HR_2<0.6})$ & 38 & 2.74$^{+0.06}_{-0.05}$ & \nodata\\
HMXB & 38 & 2.44$\pm0.05$ & \nodata \\
CV   & 18 & \nodata & 22.68$^{+2.39}_{-2.08}$ \\
SNR  & 5 &  1.84$^{+0.32}_{-0.41}$ & 6.23$^{+3.80}_{-2.67}$\\
\enddata
\tablenotetext{a}{Best-fit temperature for a black-body model (LMXBs)
or a bremmstrahlung model (CVs/SNRs)}.

\end{deluxetable}

\begin{deluxetable}{lccccccc}
\tabletypesize{\scriptsize}
\tablewidth{0pt}
\tablecaption{Broken power-law fits to the LMXB and HMXB luminosity functions}
\tablehead{
\colhead{Type} & \colhead{Lum. limit} & \colhead{Nr.} & \colhead{Faint slope} & \colhead{Break} & \colhead{Bright slope} &\colhead{Total lum.\tablenotemark{a}} &\colhead{Total luminosity\tablenotemark{b}}}
\startdata
LMXBs & $8\times10^{34}$ & 81.4$\pm15.0$ & $0.9^{+0.4}_{-0.3}$ & $3.0^{+1.8}_{-1.6}\times10^{36}$& $2.4^{+0.4}_{-0.7}$ & $1.7\pm0.4\times10^{38}$ & $1.5^{+1.5}_{-0.3}\times10^{38}$\\
HMXBs & $4\times10^{34}$ & $73.1\pm15.4$ & $1.3^{+0.2}_{-0.2}$ & $2.5^{+20}_{-2.3}\times10^{37}$ & $>2$ & $1.3\pm0.6\times10^{38}$ & $1.4^{+1.5}_{-0.3}\times10^{38}$\\
\enddata
\tablenotetext{a}{From summing individual source luminosities corrected for incompleteness.}
\tablenotetext{b}{From integrating the fitted luminosity function.}
\label{tab:fits}
\end{deluxetable}

\end{document}